\documentclass[10pt,letterpaper]{article}
\usepackage[top=0.85in,left=2.75in,footskip=0.75in]{geometry}

\usepackage{amsmath,amssymb}

\usepackage{changepage}

\usepackage[utf8x]{inputenc}

\usepackage{textcomp,marvosym}

\usepackage{cite}

\usepackage{nameref,hyperref}

\usepackage[right]{lineno}

\usepackage{microtype}
\DisableLigatures[f]{encoding = *, family = * }

\usepackage[table]{xcolor}

\usepackage{array}

\newcolumntype{+}{!{\vrule width 2pt}}

\newlength\savedwidth

\newcommand\thickhline{\noalign{\global\savedwidth\arrayrulewidth\global\arrayrulewidth 2pt}%
\hline
\noalign{\global\arrayrulewidth\savedwidth}}

\raggedright
\usepackage[aboveskip=1pt,labelfont=bf,labelsep=period,justification=raggedright,singlelinecheck=off]{caption}

\makeatletter
\renewcommand{\@biblabel}[1]{\quad#1.}
\makeatother

\usepackage{lastpage,fancyhdr,graphicx}
\usepackage{epstopdf}
\fancyheadoffset[L]{2.25in}
\fancyfootoffset[L]{2.25in}
\usepackage{soul}
\begin{document}
\vspace*{0.2in}

\begin{flushleft}
{\Large
\textbf\newline{Gait Characterization in Duchenne Muscular Dystrophy (DMD) Using a Single-Sensor Accelerometer: Classical Machine Learning and Deep Learning Approaches} 
}
\newline
\\
Albara Ah Ramli\textsuperscript{1}, 
Xin Liu\textsuperscript{1}, 
Kelly Berndt\textsuperscript{2}, 
Erica Goude\textsuperscript{2}, 
Jiahui Hou\textsuperscript{3}, 
Lynea B. Kaethler\textsuperscript{2}, 
Rex Liu\textsuperscript{1}, 
Amanda Lopez\textsuperscript{2}, 
Alina Nicorici\textsuperscript{2}, 
Corey Owens\textsuperscript{4}, 
David Rodriguez\textsuperscript{2}, 
Jane Wang\textsuperscript{2}, 
Huanle Zhang\textsuperscript{1}, 
Daniel Aranki\textsuperscript{5}, 
Craig M. McDonald\textsuperscript{2}, 
Erik K. Henricson\textsuperscript{2,6*} 
\\

\bigskip
\textbf{1} Department of Computer Science, School of Engineering; University of California, Davis, CA, USA
\\
\textbf{2} Department of Physical Medicine and Rehabilitation, School of Medicine; University of California, Davis, CA, USA
\\
\textbf{3} Department of Electrical and Computer Engineering, School of Engineering; University of Waterloo, Waterloo, Ontario, Canada
\\
\textbf{4} UC Davis Center for Health and Technology; University of California, Davis, CA, USA
\\
\textbf{5} Berkeley School of Information; University of California Berkeley, Berkeley, CA, USA
\\
\textbf{6} Graduate Group in Computer Science (GGCS); University of California, Davis, CA, USA
\\
\bigskip

%
%





* ehenricson@ucdavis.edu

\end{flushleft}
\section*{Abstract}
Differences in gait patterns of children with Duchenne muscular dystrophy (DMD) and typically-developing (TD) peers are visible to the eye, but quantifications of those differences outside of the gait laboratory have been elusive. In this work, we measured vertical, mediolateral, and anteroposterior acceleration using a waist-worn iPhone accelerometer during ambulation across a typical range of velocities. Fifteen TD and fifteen DMD children from 3-16 years of age underwent eight walking/running activities, including five 25 meters walk/run speed-calibration tests at a slow walk to running speeds (SC-L1 to SC-L5), a 6-minute walk test (6MWT), a 100 meters fast-walk/jog/run (100MRW), and a free walk (FW). For clinical anchoring purposes, participants completed a Northstar Ambulatory Assessment (NSAA). We extracted temporospatial gait clinical features (CFs) and applied multiple machine learning (ML) approaches to differentiate between DMD and TD children using extracted temporospatial gait CFs and raw data. Extracted temporospatial gait CFs showed reduced step length and a greater mediolateral component of total power (TP) consistent with shorter strides and Trendelenberg-like gait commonly observed in DMD. ML approaches using temporospatial gait CFs and raw data varied in effectiveness at differentiating between DMD and TD controls at different speeds, with an accuracy of up to 100\%. We demonstrate that by using ML with accelerometer data from a consumer-grade smartphone, we can capture DMD-associated gait characteristics in toddlers to teens.



\section{Introduction}
Patterns of gait disturbance in children with Duchenne muscular dystrophy (DMD) demonstrate biomechanical compensatory substitution to overcome strength loss and progressive joint contractures. Disease progression yields temporal and spatial changes in gait analysis metrics as described by Sutherland~\cite{REF_1}, D'Angelo~\cite{REF_2}, Heberer~\cite{REF_3}, and Gaudreault~\cite{REF_4}.  Perturbations and compensatory adaptations in gait are present from the onset of walking, are progressive, and follow a predictable pattern of increasing anterior pelvic tilt, increasing foot internal rotation and decreasing hip extension in stance phase, lateral trunk lean toward the supporting limb, increased hip flexion and hip abduction and decreased ankle dorsiflexion in the swing phase~\cite{REF_1}~\cite{REF_3}~\cite{REF_4}~\cite{REF_5}. The center of pressure at foot contact shifts laterally and anteriorly until an equinus posture at foot strike predominates~\cite{REF_1}.  These progressive adaptations lead to decreased step length and cadence during ambulation, decreased relative power of anteroposterior movement, and increased relative power of mediolateral movement with concomitant impairment of gait velocity.  

Recent advancements in novel muscle-sparing therapeutics highlight the desirability of initiating early disease modifying treatment in the toddler years, but relatively few reliable tools exist for quantitative measurement of strength, function, and mobility in this age group, underscoring an urgent need to develop new tools that include that age group and extend upward to the limits of ambulation~\cite{REF_5}.  
Wearable accelerometers can accurately measure variation in step rates in children with DMD in both the laboratory and community settings, produce natural history data that is suitable for analysis in clinical trials~\cite{REF_6}, and hold promise to provide a complete picture of the effect of strength limitation on community mobility and daily activities. To maximize their effectiveness, wearable devices will need to detect and record well-understood quantitative temporal and spatial features of gait patterns while being unobtrusive and affordable.  

The increasing availability of high-quality, low-cost triaxial accelerometers and inertial measurement units as stand-alone devices or integrated into commonly-available smartphones yields new opportunities to gather community level data across a wide range of typically-developing (TD) individuals and those affected by movement-related disorders.  Because of this, researchers are developing a better understanding of how to extract and interpret temporal and spatial features of single accelerometer data that include not only step counts and frequencies but also a wide variety of other features~\cite{REF_7}~\cite{REF_13} including step lengths, step velocities~\cite{REF_4}~\cite{REF_8} and triaxial power spectra~\cite{REF_9}~\cite{REF_10} in order to use those features in principal components analysis (PCA) to evaluate between-group differences and changes over time~\cite{REF_11}. The utility of these measures in describing disease severity and tracking disease progression has been demonstrated in the golden retriever muscular dystrophy (GRMD) of DMD~\cite{dog}, as well as in children with DMD~\cite{REF_12}.

Because of the increasing availability of such sensing data, there is a strong demand for automated systems that can thoroughly analyze and utilize such data. In this project, we take the first step towards this goal. We evaluate the utility of various classical machine learning (CML) and deep learning (DL)-based approaches to differentiate between children with and without DMD using data from consumer-level mobile phone accelerometers during walking and running activities.  We developed a system (Walk4Me~\cite{dmdpropose20davis}~\cite{walk4me_page}) consisting of a smartphone-based application to collect raw data remotely using the phone's built-in accelerometer sensor, combined with a web-based tool to aggregate, store and analyze data.  We extracted the temporal/spatial gait characteristics and used CML and DL techniques~\cite{DL_REF} to evaluate the gait changes associated with DMD, using both extracted features and raw data.

We note that our project is only a first step towards an automated disease monitoring system that can be used for disease prescreening, diagnosis assistance, progression monitoring, and possibly in a subject's natural environment (which we call community-based) instead of motion labs. It should be noted DMD needs to be diagnosed in specialty clinics, which are sparsely located across the country, and a potential patient could be hundreds of miles away from the closest clinic. The tool described in this work can be used to facilitate prescreening for such patients. Additionally, this tool, further developed, could be used to continue monitoring the progression of the disease and quantify the effectiveness of medical treatment. Furthermore, while our current system and work focus on DMD, it has the potential to expand to other mobility-related diseases, such as post-stroke recovery and healthy aging.

\section{Materials and Methods}
\subsection{Participants}
The University of California, Davis institutional review board (IRB) reviewed and approved the protocol. Informed consent was obtained for each participant prior to initiation of study procedures. We studied thirty male participants (15 with DMD, 15 TD) who were between the ages of 3 and 16 years old, had at least 6 months of walking experience, and could perform a 10 meters walk/jog/run test in less than 10 seconds. Participants with DMD had a confirmed clinical diagnosis and were glucocorticoid therapy-naïve or on a stable regimen for at least three months. Northstar Ambulatory Assessment (NSAA) scores for DMD participants ranged from 34 to 8, indicating typical levels of function to clinically-apparent moderate mobility limitation (Table-\ref{table1}). 

\begin{table}[!ht]
\begin{adjustwidth}{0in}{0in} 
\caption{\bf Characteristic of the children included in the study.}\label{table1}
\begin{tabular}{|p{1.4cm}|p{1.4cm}|p{1.4cm}|p{1.4cm}|p{1.4cm}|p{1.4cm}|} 
\thickhline
\textbf{Case} & \textbf{Value} & \textbf{Age} & \textbf{Weight} & \textbf{Height} & \textbf{NSAA}\\ 
 (status)  &  & (years) & (kg) & (cm) & (/34)\\\thickhline
 & Mean & 9.5 & 37.7 & 127.1 & 20.5\\ 
\cline{2-6}
\textbf{DMD} & (SD) & (3.9) & (16.0) & (16.2) & (8.2)\\ 
\cline{2-6}
(N=15) & Min & 3 & 17.2 & 101.6 & 8.0\\ 
\cline{2-6}
 & Max & 16 & 67.7 & 153.3 & 34.0\\ 
\cline{2-6}
\thickhline
 & Mean & 7.7 & 34.2 & 130.8 & 33.8\\ 
\cline{2-6}
\textbf{TD} & (SD) & (3.0) & (21.6) & (15.8) & (0.8)\\ 
\cline{2-6}
(N=15) & Min & 4 & 18.6 & 108.5 & 31.0\\ 
\cline{2-6}
 & Max & 15 & 101.0 & 165.5 & 34.0\\ 
\cline{2-6}
\thickhline
\textbf{} & p-value & 0.1664 & 0.6229 & 0.5331 & \textbf{$<$0.0001}\\ 
\cline{2-6}
\thickhline
\end{tabular}
\end{adjustwidth}
\end{table}
Scores for TD participants ranged from 34 to 31, indicating a maximal range of task performance, with a low score of 31 in a six-year-old participant indicating age-appropriate achievement of developmental milestones~\cite{REF_MERCURI_NSAA}.

\subsection{Materials}
\begin{figure}[!ht]
\centerline{\includegraphics[height=1.3in]{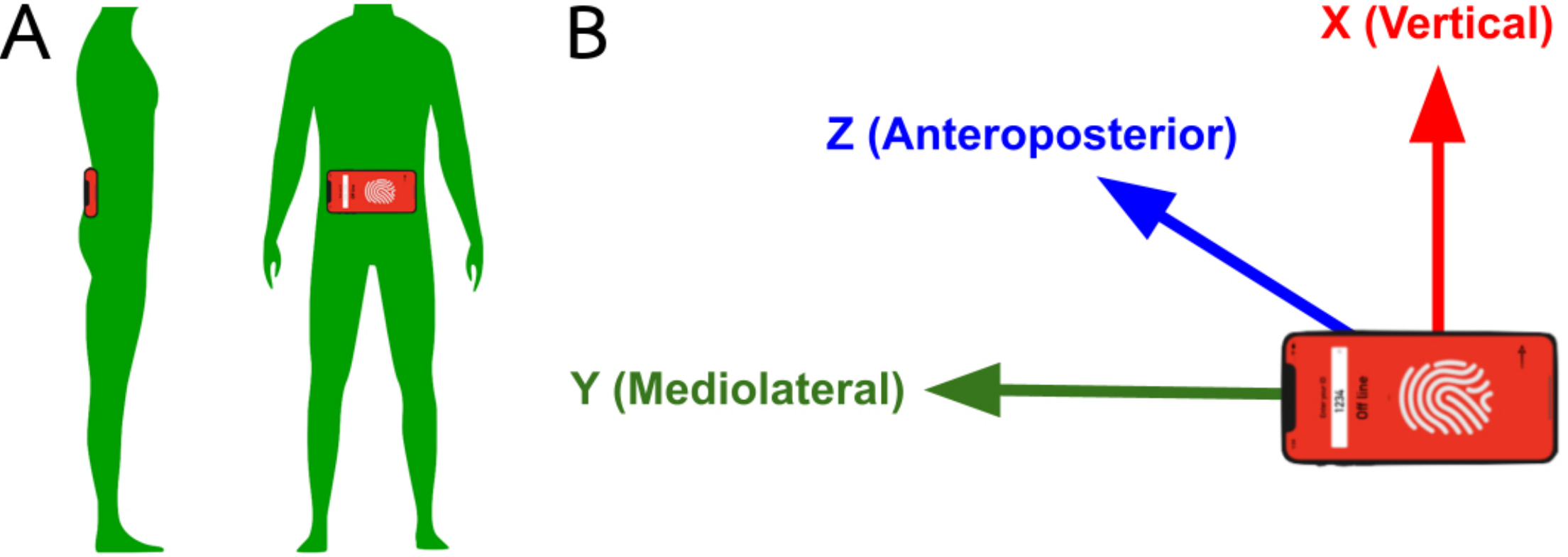}}
\caption{{\bf The position and the orientation of the accelerometer.}
(A) The smartphone with its built-in accelerometer is positioned near the body's center of mass, fastened by a belt around the waist. (B) The accelerometer records acceleration in three axes. The x-axis is vertical, the y-axis is mediolateral, and the z-axis is anteroposterior.}
\label{Fig1}
\end{figure}

Participants wore an NGN Sport fitness phone belt (Engine Design Group, LLC, Los Angeles, CA) carrying an Apple iPhone 11 (Apple, Inc. Cupertino, CA), which includes a single, built-in triaxial accelerometer.  The belt was placed close to the body’s center of mass at the lumbosacral junction at the approximate level of the iliac crest, with the phone oriented so that the right lower side was consistently oriented in the upper right position to measure acceleration in the vertical, mediolateral and anteroposterior axes (Fig~\ref{Fig1}A and Fig~\ref{Fig1}B). We developed an iPhone app (Walk4Me) to continuously stream raw sensor data to a cloud server at a sampling rate of 100 Hz.  At the conclusion of each assessment session, data was processed via Walk4Me's web application to extract clinical gait features and train the CML and DL models. 

\subsection{Gait and Functional Testing}
\label{sec:gait-analysis}
Children performed eight walking and running activities along a 25 meters straight-line course according to established protocols~\cite{REF_McDonald_Henricson_6MWT}. The first five activities were speed-calibration-tests (SC) where the child walked along a 25 meters long corridor at incrementally increasing gait velocities every 25 meters from slowest speed, speed-calibration-L1 (SC-L1), to a slow-walk (SC-L2), to a comfortable self-selected-walk pace (SC-L3), to a fast-walk (SC-L4), to a run or fastest possible gait (SC-L5)~\cite{REF_4}. The evaluator recorded the observed number of steps taken during each effort.  Participants then performed a 6-minute walk test (6MWT), followed by a 100 meters fast-walk/jog/run (100MRW) using previously described methods~\cite{REF_McDonald_Henricson_6MWT}~\cite{alfano_2017}.  The evaluator recorded the 6MWT distance and the number of steps in the first 50 meters, and the time to complete the 100MRW. 

Participants also completed the NSAA according to established protocols~\cite{mazzone2009reliability}.  The NSAA consists of a set of 17 graded tasks representing common mobility and positional transfer activities.  Tasks are graded individually with a total score of 34 points indicating full functional mobility.  All evaluations were recorded on video for later verification of step counts, distances, task times, and task quality.

\subsection{Data Analysis}
\begin{figure}[!ht]
\centerline{\includegraphics[width=\textwidth]{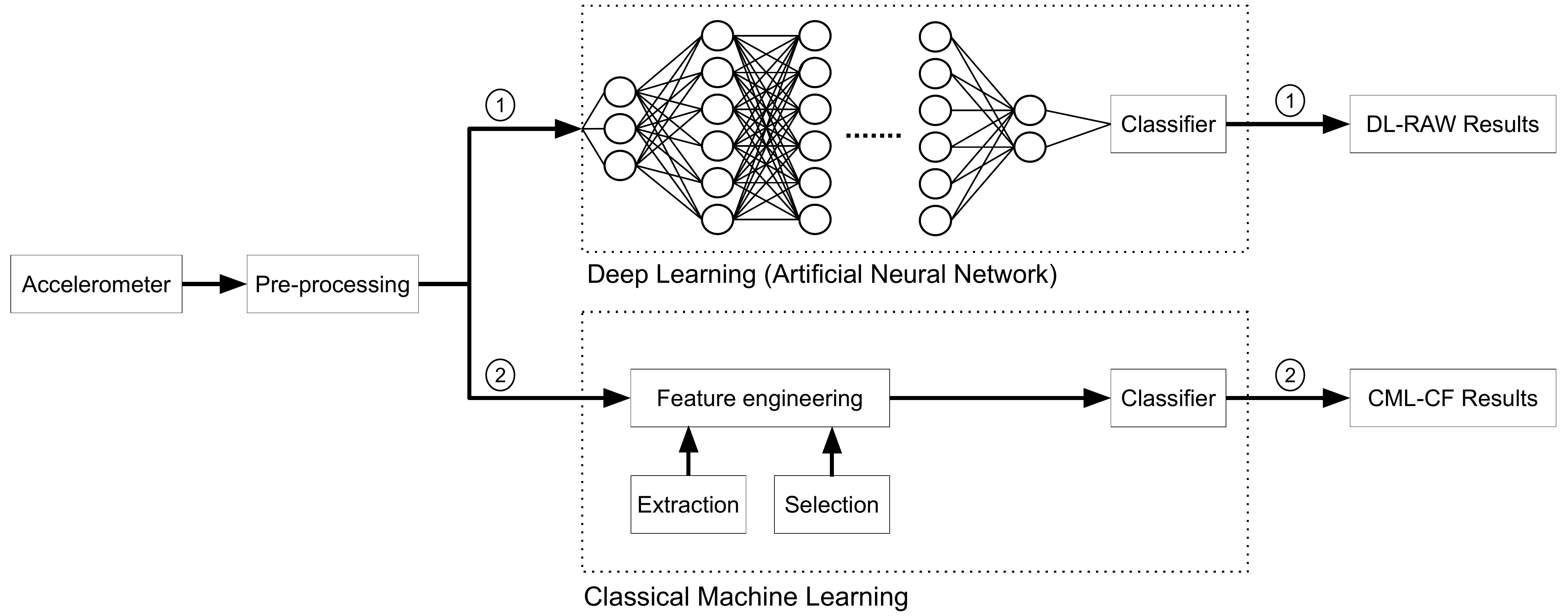}}
\caption{{\bf A typical process diagram of classical machine learning (CML-CF) and deep learning (DL-RAW).}
The two methods used are (1) DL-RAW and (2) CML-CF.}
\label{Fig2}
\end{figure}

Data analysis was performed using both CML and DL approaches (Fig~\ref{Fig2}).  In our first approach, we processed raw data from each activity to extract temporospatial gait CFs including speed, step length, step frequency, total power, percent of power in each axis, and force index in a manner similar to those described by Barthelemy~\cite{dog} and Fraysse~\cite{dogLDA}. We normalized speed and step length to standing height. For information purposes, we also compared the means of features between DMD and TD control groups for each activity using simple two-tailed T-tests. For convenience, we bold the p-values when it is smaller than 0.05, without a Bonferroni adjustment.
We note p-values are reported here primarily to describe relationships between temporospatial gait CFs and model dimensions across DMD and control groups - they are not considered in ML algorithms. A feature with a non-statistically-significant p-value could still be useful in ML tasks. This is especially true when the sample size is relatively small, as in this study.
We applied CML approaches to the temporospatial gait CFs both with and without dimensional reduction to identify data characteristics and models that were most accurate at predicting group membership.  We used Pearson's Correlation Coefficients with a Bonferroni adjustment to describe relationships between temporospatial gait CFs and latent domain scores after dimensional reduction, except during comparisons with ordinal measures where we used Spearman's Rank Ordered Correlation.

\subsubsection{Extraction and Evaluation of Temporospatial Gait Clinical Features}
We extracted the following eight temporospatial gait CFs from the raw accelerometer data:\\
\textbf{Speed (SP)} is measured in meters per second and normalized by the height of the child in meters. This feature is calculated by dividing the distance by the time spent performing each activity.\\

\begin{figure}[!ht]
\centerline{\includegraphics[width=1\textwidth]{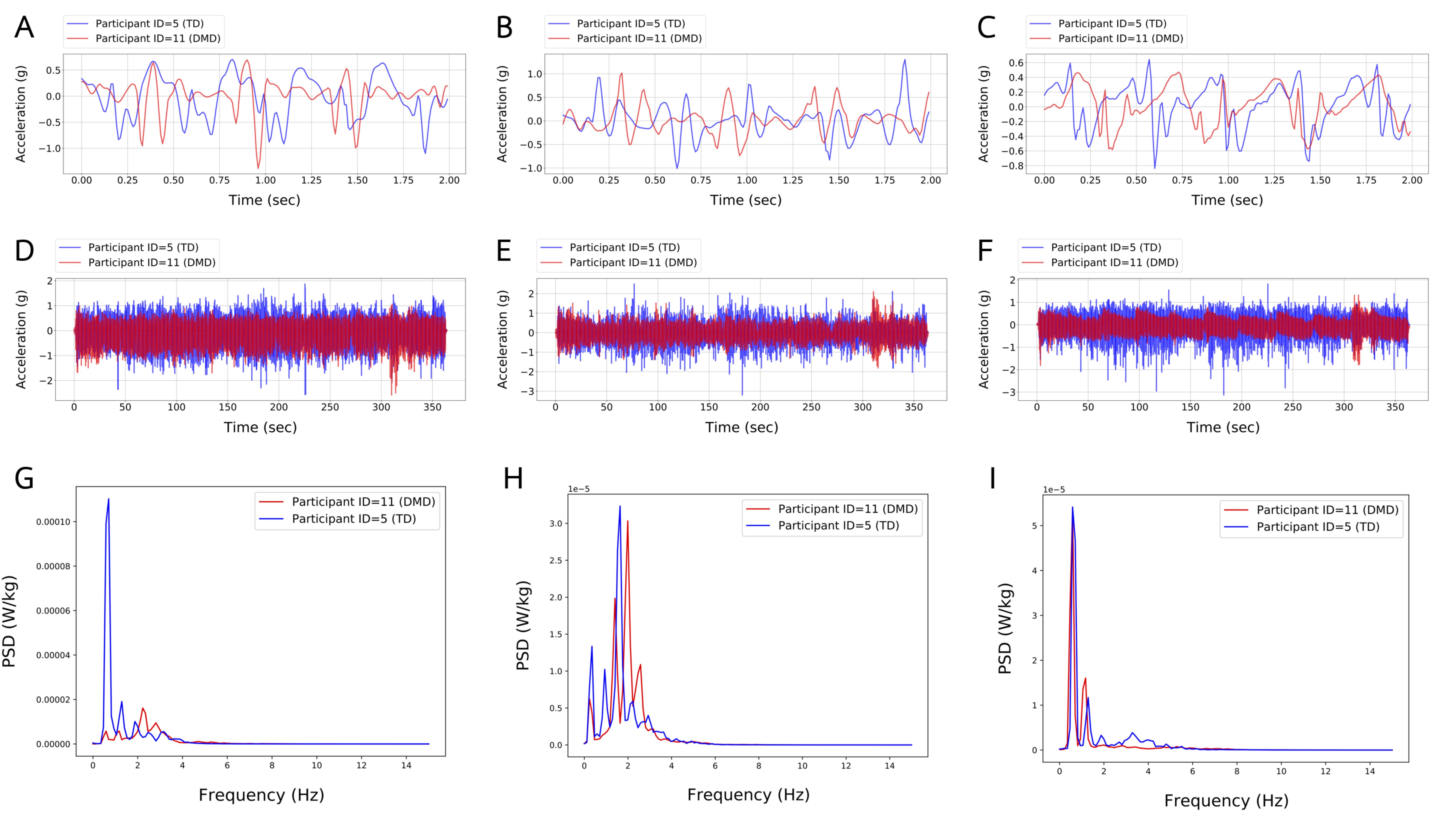}}
\caption{{\bf Figures Illustrate a comparison of two children, DMD (ID=11) and TD (ID=5), in the three-axial accelerations Vertical (x-axis), Mediolateral (y-axis), and Anteroposterior (z-axis) of 6MWT.}
Figures (A, B, and C) represent a 2 second TW acceleration signal of a TD child performing 5 steps, and a DMD child performing 4 steps. Figures (D, E, and F) represent the acceleration signal of a TD child and a DMD child performing the whole 6MWT. Figures (G, H, and I) represent the Power-Spectrum-Density (PSD) of a TD child and a DMD child performing the whole 6MWT.}
\label{Fig3}
\end{figure}

\textbf{Step Length (SL)} is measured in meters and normalized by the height of the child in meters. This feature is calculated by dividing the total distance by the number of steps. The number of steps is calculated using a low pass filter on the anteroposterior (z-axis) acceleration signal (Fig~\ref{Fig3}C and Fig~\ref{Fig3}F). Then, we calculate the number of peaks, where each peak represents one step, and a complete gait cycle is composed of two steps.\\
\textbf{Step Frequency (SF)} is measured in steps per second. This feature is calculated by dividing the number of steps by the time for each activity. \\

\textbf{Total Power (TP)} is measured in W/kg. This feature is calculated by first transferring the time domain to the frequency domain of the three axes using Fast Fourier Transform (FFT): vertical (x-axis), mediolateral (y-axis), and anteroposterior (z-axis) (Fig~\ref{Fig3}D, Fig~\ref{Fig3}E, and Fig~\ref{Fig3}F, respectively). Then, we sum the integral of the power (normalized by weight) in each of the three axes (Fig~\ref{Fig3}G, Fig~\ref{Fig3}H, and Fig~\ref{Fig3}I).\\

\textbf{Mediolateral Power (MP)} is measured in \% of TP. This feature is calculated by first transferring the time domain of the accelerometer's y-axis (Fig~\ref{Fig3}E) to the frequency domain using FFT. Then we calculate the integral of the power spectrum density (PSD) (Fig~\ref{Fig3}H). Finally, we normalize the value by weight and TP.\\

\textbf{Anteroposterior Power (AP)} is measured in \% of TP. This feature is calculated by first transferring the time domain of the accelerometer's z-axis (Fig~\ref{Fig3}F) to the frequency domain using FFT. Then we calculate the integral of the PSD (Fig~\ref{Fig3}I). Finally, we normalize the value by weight and TP.\\

\textbf{Vertical Power (VP)} is measured in \% of TP. This feature is calculated by first transferring the time domain of the accelerometer's x-axis (Fig~\ref{Fig3}D) to the frequency domain using FFT. Then we calculate the integral of the PSD (Fig~\ref{Fig3}G). Finally, we normalize the value by weight and TP.\\

\textbf{Force Index (FI)} is measured in N/kg. This feature is calculated by first transferring the time domain of the accelerometer's z-axis (Fig~\ref{Fig3}F) to the frequency domain using FFT. Then, we divide the integral of the PSD (Fig~\ref{Fig3}I) by the average speed in order to average the force index.

\subsubsection{CML and DL Analytical Methods} 
\begin{figure}[!ht]
\centerline{\includegraphics[width=1\textwidth]{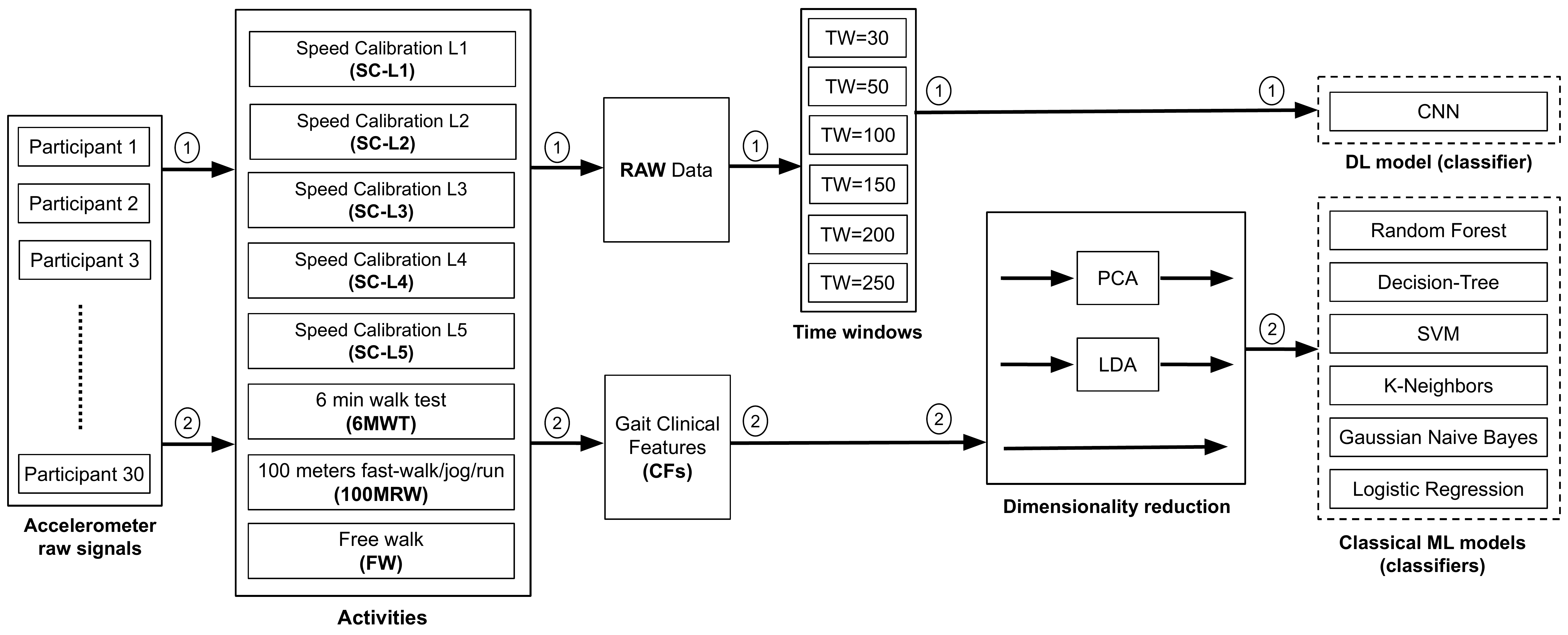}}
\caption{{\bf The process diagram of (1) the DL-RAW approach and (2) the CML-CF approach.}
}
\label{Fig4}
\end{figure}

We used two different ML approaches to classify accelerometer data collected at a range of gait speeds as belonging to a child with DMD or a TD.  For the \textbf{CML-CF} approach, 6 different classifiers were implemented: Random Forest (RF), Decision-Tree (DT), Support-Vector Machine (SVM), K-Nearest Neighbors (KNN), Gaussian Naive Bayes (GNB), Logistic Regression (LR). In this method, we used the eight extracted temporospatial gait CFs as input into CML classifiers with and without dimensionality reduction. The eight extracted temporospatial gait CFs for each child were used to train the CML model for all eight activities. For the \textbf{DL-RAW} method, we used a Convolutional Neural Network (CNN) model with the time-windowed raw accelerometer signal as an input in the DL classifier. Fig~\ref{Fig4} shows the process diagram  of the DL-RAW and CML-CF approaches. 

\subsubsection{Dimensionality Reduction}
To analyze how the participants are distributed on the projection plane, we used dimensionality reduction techniques in a manner similar to those described by Fraysse and Barthelemy~\cite{dogPCA}~\cite{dogLDA}. We investigated whether using principal component analysis (PCA) and linear discriminant analysis (LDA) techniques would affect the discrimination accuracy between DMD and TD groups. We fed the eight extracted temporospatial gait CFs to CML models using PCA and LDA. We used PCA and LDA to reduce the dimensionality of the input  features of all participants and projected their models' results into a two-dimensional (2D) and one-dimensional (1D) representation, respectively.  We compared PCA and LDA accuracy with the original models without using any dimensionality reduction techniques. 

\subsubsection{Preprocessing of Raw Accelerometer Signals Using Time-windowing}
\label{sec:TW}
\begin{figure}[!ht]
\centerline{\includegraphics[scale=.3]{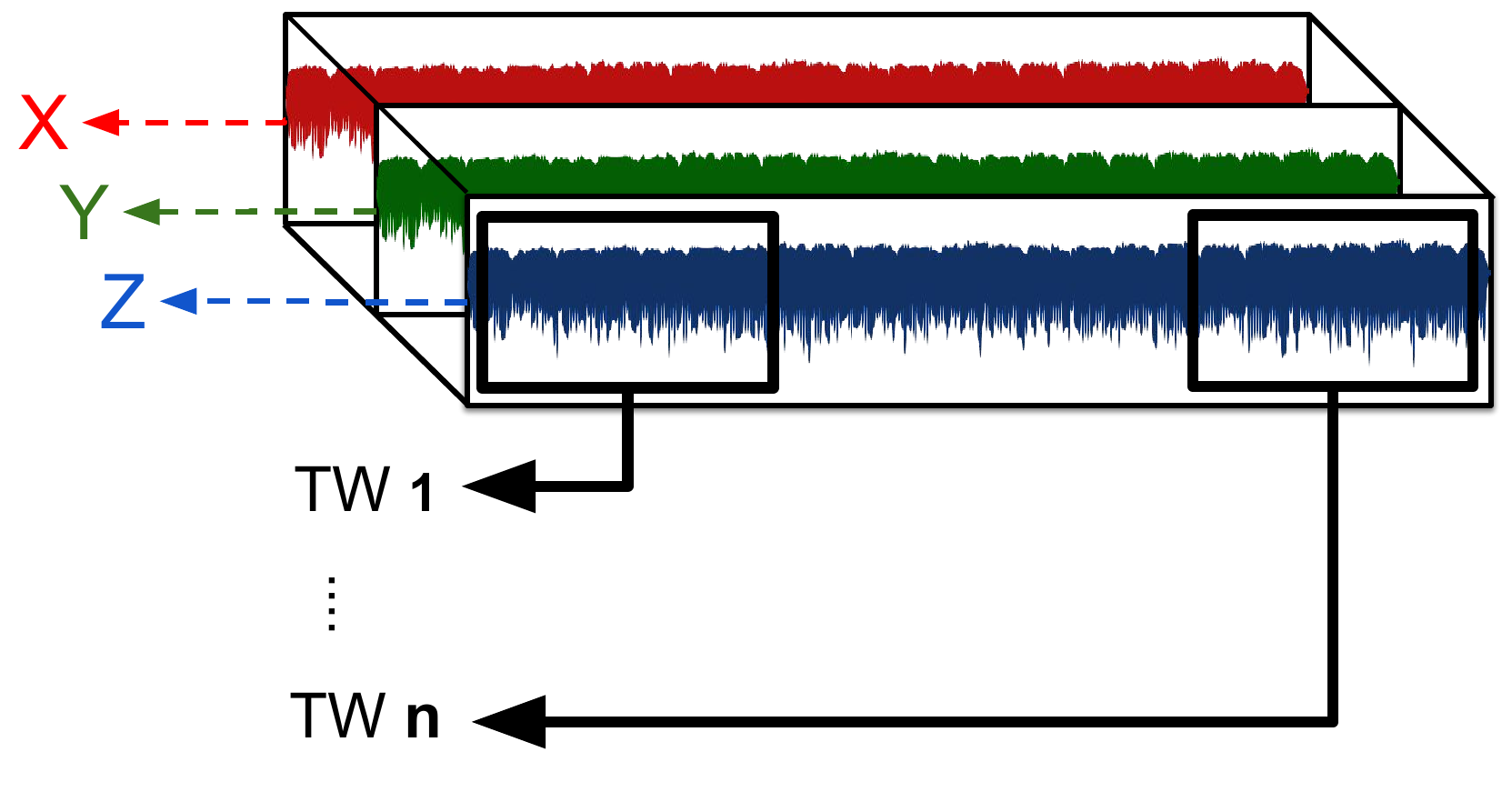}}
\caption{{\bf Time-windows (TWs) of the accelerometer (x, y, and z axes)}
}
\label{Fig5}
\end{figure}

In the CML-CF method, data from each activity for each participant represents an individual input to the model. The temporospatial gait CFs of each activity must be extracted entirely before using it as input to the model. In the DL-RAW method, we used raw acceleration values in each of the three axes as input to the DL model (Fig~\ref{Fig4}). We used the window-slicing method~\cite{cui2016multi}~\cite{review} to segment the raw data from each activity into multiple fixed time-windows (TWs) (Fig~\ref{Fig5}) in order to augment the data input~\cite{le2016data} and to facilitate DL model convergence~\cite{bwcnn}~\cite{REF_CNN}. We examined six distinct TWs (i.e., 0.3, 0.5, 1, 1.5, 2, 2.5 seconds) to determine the signal duration required for the highest model accuracy. For each activity, the model predicted whether an individual TW was labeled as DMD or TD. We then used those predictions to calculate the overall percentage of the TWs predicted as DMD or TD. At inference time, we examined the percentage of typical/atypical decisions for each TW, and used majority voting to determine whether the participants were correctly labeled as having DMD.  

\subsubsection{Cross-Validation of ML Models}
We used Leave-One-Subject-Out (LOSO) cross-validation to evaluate classification accuracy in the presence of slight variations in dataset contents. We use LOSO to ensure that the model does not have information leakage. Having data from a particular participant in both training and testing datasets makes the model more familiar with the data of that participant. Thus, using a completely unseen participant in the testing-set will ensure that the model is not biased toward that participant because it has never been trained on any part of that participant's data before. The present dataset included 30 participants. To ensure the accuracy of our predictions, we utilized data from 29 of the participants as the training-set and predicted the label of the remaining participant. We repeated this process 30 times, each time leaving out a different participant for testing, and calculated the accuracy of each model by averaging the 30-fold accuracy. By using this method, we were able to ensure a robust and reliable analysis of our data.

\section{Results}

\subsection{Extracted Temporospatial Gait Clinical Features}
We compared temporospatial gait CFs of DMD and TD participants extracted from our sensor data, shown in Table-\ref{table2}. For the 25 meters course at a slow walking pace (SC-L1 and SC-L2), walking speed and step length were significantly lower, and the mediolateral percent of accelerations was significantly higher in people with DMD. Those observations are consistent with clinical descriptions of a slower and more lateral DMD gait. On our 25 meters and more extended efforts, self-selected-walk and fast-walk paces (SC-L3, FW, SC-L4, and 6MWT), we saw a similar pattern of differences that added a lower percent of vertical accelerations in people with DMD. At jogging to running paces (SC-L5 and 100MRW), significantly reduced step frequency of people with DMD became apparent. When combining all efforts, differences in speed, step frequency, step length, total power, and percent power in vertical and mediolateral axes differed significantly between the two groups. Interestingly, there was little between-group-difference in anteroposterior force at any but the slowest of velocities.

\begin{table}[!ht]
\begin{adjustwidth}{-2.25in}{0in} 
\centering
\caption{\bf The summary of temporospatial gait CFs for the eight different gait activities.}
\label{table2}
\resizebox{1.0\textwidth}{!}{
\begin{tabular}{|p{1.8cm}||p{0.8cm}|p{1.3cm}||p{1.2cm}|p{1cm}|p{1.2cm}|p{1.2cm}|p{1.2cm}|p{1.2cm}|p{1cm}|p{1.2cm}|} 
\thickhline
\multicolumn{1}{|c||}{\textbf{}} & \multicolumn{1}{c|}{\textbf{}} & \multicolumn{1}{c||}{\textbf{}} & \multicolumn{8}{c|}{\textbf{Temporospatial Gait Clinical Features (CFs)}} \\  
\cline{4-11}
\textbf{Activities} & \textbf{Case} & \textbf{Value} & \textbf{SP \textbf{ } $(\%)$} & \textbf{SF} & \textbf{SL \textbf{ } $(\%)$} & \textbf{TP $(10^{-6})$} & \textbf{VP  \textbf{ } $(\%)$} & \textbf{MP $(\%)$} & \textbf{AP \textbf{ } $(\%)$} & \textbf{FI $(10^{-3})$} \\ 
\thickhline
 & TD & Mean & 0.35 & 1.29 & 0.27 & 64.84 & 28.45 & 36.24 & 35.31 & 11.54\\ 
 &  & (SD) & (0.06) & (0.19) & (0.02) & (48.52) & (4.07) & (7.22) & (5.45) & (8.94)\\ 
\cline{2-11}\textbf{SC-L1} & DMD & Mean & 0.26 & 1.12 & 0.23 & 62.0 & 27.82 & 42.1 & 30.08 & 5.79\\ 
 &  & (SD) & (0.1) & (0.32) & (0.04) & (92.77) & (5.48) & (6.17) & (4.36) & (6.95)\\ 
\cline{2-11} &  & p-value & \textbf{0.0077} & 0.093 & \textbf{0.0004} & 0.9172 & 0.7246 & \textbf{0.0239} & \textbf{0.0072} & 0.0594\\ 
\thickhline
 & TD & Mean & 0.55 & 1.62 & 0.34 & 155.3 & 31.48 & 34.29 & 34.23 & 16.02\\ 
 &  & (SD) & (0.07) & (0.16) & (0.02) & (104.07) & (7.78) & (8.4) & (5.63) & (13.4)\\ 
\cline{2-11}\textbf{SC-L2} & DMD & Mean & 0.46 & 1.58 & 0.29 & 170.27 & 28.25 & 40.16 & 31.59 & 9.85\\ 
 &  & (SD) & (0.12) & (0.31) & (0.03) & (179.9) & (6.2) & (6.8) & (6.63) & (11.03)\\ 
\cline{2-11} &  & p-value & \textbf{0.0139} & 0.6133 & \textbf{$<$0.0001} & 0.7823 & 0.22 & \textbf{0.0445} & 0.2487 & 0.1801\\ 
\thickhline
 & TD & Mean & 0.76 & 1.89 & 0.4 & 355.84 & 36.93 & 31.36 & 31.7 & 19.54\\ 
 &  & (SD) & (0.12) & (0.23) & (0.02) & (255.55) & (10.22) & (7.9) & (5.72) & (16.06)\\ 
\cline{2-11}\textbf{SC-L3} & DMD & Mean & 0.63 & 1.89 & 0.33 & 334.35 & 30.16 & 39.32 & 30.52 & 14.05\\ 
 &  & (SD) & (0.1) & (0.24) & (0.03) & (317.31) & (5.98) & (7.48) & (5.79) & (13.76)\\ 
\cline{2-11} &  & p-value & \textbf{0.0024} & 0.9457 & \textbf{$<$0.0001} & 0.8396 & \textbf{0.0348} & \textbf{0.0085} & 0.5786 & 0.323\\ 
\thickhline
 & TD & Mean & 1.28 & 2.55 & 0.5 & 1907.94 & 38.92 & 30.94 & 30.14 & 58.39\\ 
 &  & (SD) & (0.24) & (0.32) & (0.04) & (1690.83) & (8.75) & (8.73) & (5.84) & (58.04)\\ 
\cline{2-11}\textbf{SC-L4} & DMD & Mean & 0.94 & 2.4 & 0.39 & 1807.78 & 29.07 & 40.11 & 30.83 & 49.58\\ 
 &  & (SD) & (0.28) & (0.51) & (0.06) & (3060.77) & (10.71) & (9.29) & (3.86) & (65.31)\\ 
\cline{2-11} &  & p-value & \textbf{0.0016} & 0.3496 & \textbf{$<$0.0001} & 0.9125 & \textbf{0.0101} & \textbf{0.0095} & 0.7091 & 0.6992\\ 
\thickhline
 & TD & Mean & 2.44 & 3.61 & 0.67 & 9219.81 & 50.13 & 21.53 & 28.34 & 93.53\\ 
 &  & (SD) & (0.48) & (0.52) & (0.09) & (6775.47) & (8.93) & (10.08) & (10.09) & (71.93)\\ 
\cline{2-11}\textbf{SC-L5} & DMD & Mean & 1.22 & 2.82 & 0.42 & 4235.65 & 35.67 & 35.01 & 29.32 & 68.55\\ 
 &  & (SD) & (0.48) & (0.73) & (0.07) & (5913.64) & (13.68) & (10.86) & (9.18) & (88.33)\\ 
\cline{2-11} &  & p-value & \textbf{$<$0.0001} & \textbf{0.002} & \textbf{$<$0.0001} & \textbf{0.0406} & \textbf{0.0019} & \textbf{0.0015} & 0.7822 & 0.4029\\ 
\thickhline
 & TD & Mean & 1.18 & 2.36 & 0.5 & 1521.86 & 41.68 & 29.51 & 28.81 & 915.74\\ 
 &  & (SD) & (0.11) & (0.17) & (0.03) & (1054.76) & (11.07) & (9.01) & (6.71) & (677.16)\\ 
\cline{2-11}\textbf{6MWT} & DMD & Mean & 0.79 & 2.05 & 0.38 & 992.04 & 31.61 & 38.79 & 29.6 & 840.77\\ 
 &  & (SD) & (0.25) & (0.39) & (0.06) & (1441.95) & (6.02) & (6.93) & (5.46) & (1210.56)\\ 
\cline{2-11} &  & p-value & \textbf{$<$0.0001} & \textbf{0.0088} & \textbf{$<$0.0001} & 0.2664 & \textbf{0.0056} & \textbf{0.0046} & 0.7327 & 0.837\\ 
\thickhline
 & TD & Mean & 2.27 & 3.39 & 0.67 & 8402.12 & 52.89 & 18.05 & 29.06 & 480.5\\ 
 &  & (SD) & (0.4) & (0.47) & (0.08) & (5348.93) & (8.86) & (9.36) & (10.53) & (410.11)\\ 
\cline{2-11}\textbf{100MRW} & DMD & Mean & 1.1 & 2.57 & 0.42 & 2944.45 & 39.68 & 34.82 & 25.51 & 599.57\\ 
 &  & (SD) & (0.42) & (0.67) & (0.07) & (5073.6) & (17.19) & (12.28) & (7.58) & (955.96)\\ 
\cline{2-11} &  & p-value & \textbf{$<$0.0001} & \textbf{0.0013} & \textbf{$<$0.0001} & \textbf{0.0148} & \textbf{0.0172} & \textbf{0.0006} & 0.3512 & 0.6685\\ 
\thickhline
 & TD & Mean & 0.83 & 1.96 & 0.42 & 617.5 & 41.38 & 28.46 & 30.17 & 321.34\\ 
 &  & (SD) & (0.16) & (0.29) & (0.04) & (478.61) & (8.92) & (7.49) & (4.87) & (263.17)\\ 
\cline{2-11}\textbf{FW} & DMD & Mean & 0.61 & 1.83 & 0.33 & 529.12 & 32.87 & 36.47 & 30.66 & 638.46\\ 
 &  & (SD) & (0.18) & (0.42) & (0.06) & (983.01) & (8.51) & (6.99) & (7.19) & (1252.37)\\ 
\cline{2-11} &  & p-value & \textbf{0.0015} & 0.3626 & \textbf{$<$0.0001} & 0.7565 & \textbf{0.0124} & \textbf{0.0052} & 0.826 & 0.3454\\ 
\thickhline
 & TD & Mean & 1.21 & 2.33 & 0.47 & 2780.65 & 40.23 & 28.8 & 30.97 & 239.57\\ 
 &  & (SD) & (0.76) & (0.83) & (0.14) & (4680.29) & (11.6) & (10.13) & (7.35) & (417.76)\\ 
\cline{2-11}\textbf{All} & DMD & Mean & 0.74 & 2.01 & 0.35 & 1333.61 & 31.62 & 38.47 & 29.91 & 262.26\\ 
 &  & (SD) & (0.4) & (0.69) & (0.08) & (3175.88) & (10.1) & (8.54) & (6.43) & (740.37)\\ 
\cline{2-11} &  & p-value & \textbf{$<$0.0001} & \textbf{0.0016} & \textbf{$<$0.0001} & \textbf{0.0062} & \textbf{$<$0.0001} & \textbf{$<$0.0001} & 0.2423 & 0.7714\\ 
\thickhline
\end{tabular}}
\begin{flushleft}
The level of significance was set at 0.05. SP: speed in meters per second normalized by height in meters. TP: total Power with a unit of $10^{-6}$ W/kg. SF: Step Frequency with a unit of step/second. SL: Step length as a percentage of standing height. MP: the percentage of power in the x-axis normalized by total power TP (\%). AP: the percentage of power in the y-axis normalized by total power TP (\%). VP: the percentage of power in the z-axis normalized by total power TP (\%). FI: force index with a unit of $10^{-3}$ N/kg. 
\end{flushleft}
\end{adjustwidth}
\end{table}

\subsection{Comparison between CML-CF and DL-RAW Approaches}
\begin{figure}[!ht]
\centerline{\includegraphics[width=4in]{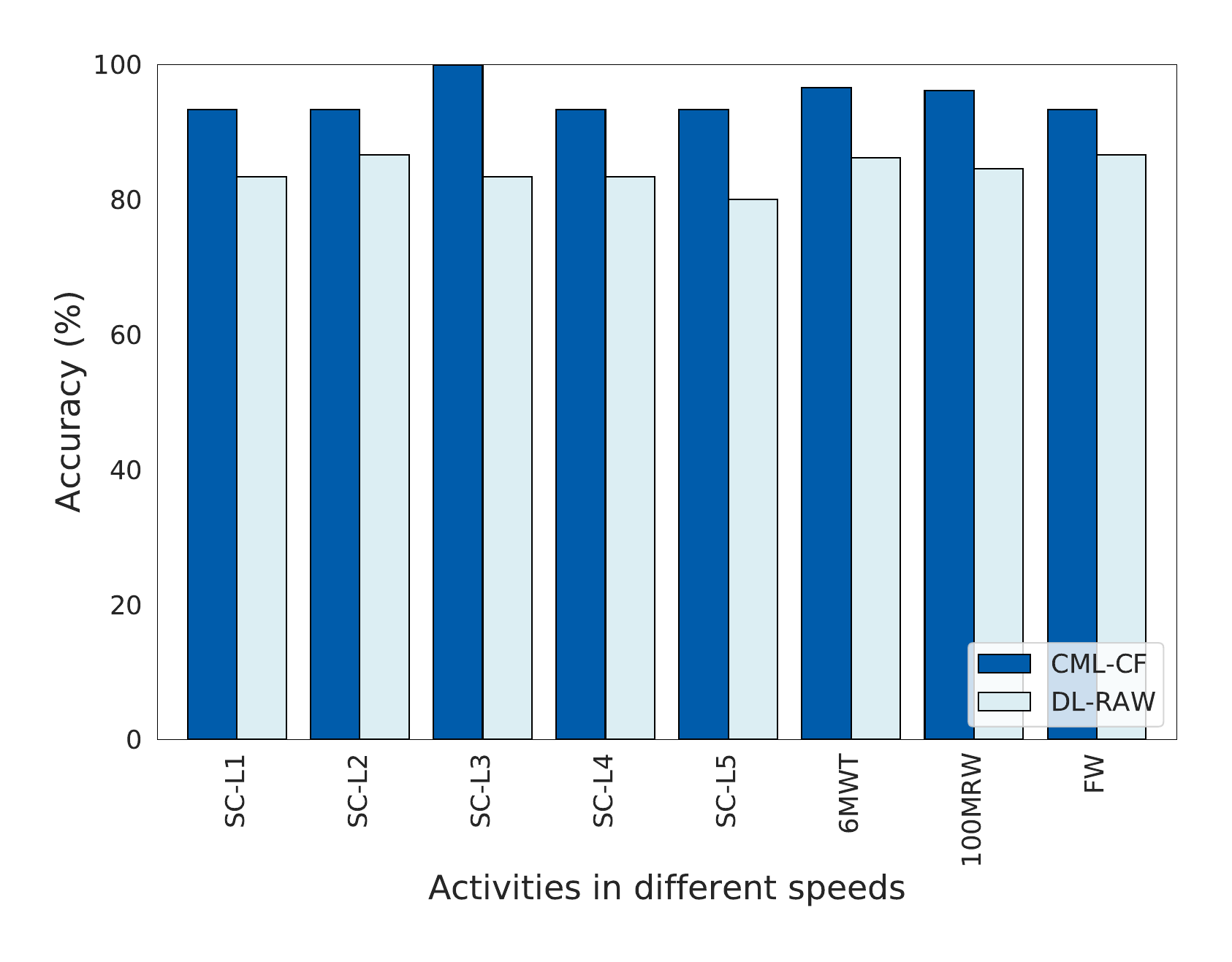}}
\caption{{\bf The optimal classifier's accuracy of CML-CF and DL-RAW.}
The y-axis represents the classifier accuracies, while the x-axis represents different activities.
}
\label{Fig6}
\end{figure}

Two different ML approaches were performed: CML-CF and DL-RAW. Fig~\ref{Fig6} and Table-\ref{table3} summarize the best results obtained from each approach. Fig~\ref{Fig6} represents the optimal accuracy obtained by performing the eight different gait activities: SC-L1, SC-L2, SC-L3, SC-L4, SC-L5, 6MWT, 100MRW, and FW (described in Section \ref{sec:gait-analysis}). LOSO was used to verify the accuracy results.

\begin{table}[!ht]
\begin{adjustwidth}{-2.25in}{0in} 
\centering
\caption{\bf Classification performance of CML-CF and DL-RAW approaches.}
\label{table3}
\resizebox{0.8\textwidth}{!}{
\begin{tabular}{|p{2cm}|p{1.5cm}|p{1.5cm}|p{1.5cm}|p{1.5cm}|p{1.5cm}|p{1.5cm}|}
\thickhline
& \multicolumn{4}{c|}{\textbf{CML-CF}} &  \multicolumn{2}{c|}{\textbf{DL-RAW}} \\  
\cline{2-7}
\textbf{Activities} & \multicolumn{1}{c|}{\textbf{Alg.}} & \multicolumn{1}{c|}{\textbf{CML}} & \multicolumn{1}{c|}{\textbf{PCA}} & \multicolumn{1}{c|}{\textbf{LDA}} & \multicolumn{1}{c|}{\textbf{TW$^{*}$}} & \multicolumn{1}{c|}{\textbf{CNN}} \\
&  & \multicolumn{1}{c|}{$(\%)$} & \multicolumn{1}{c|}{$(\%)$} & \multicolumn{1}{c|}{$(\%)$} &  \multicolumn{1}{c|}{(Samples)} & \multicolumn{1}{c|}{$(\%)$} \\ 

\thickhline
 & RF & 76.67 & \textbf{86.67} & \textbf{93.33} & 30 & 79.98 \\
\cline{2-7}
 & DT & 66.67 & 80.0 & \textbf{93.33} & 50 & \textbf{83.35} \\
\cline{2-7}
\textbf{SC-L1} & SVM & 73.33 & 73.33 & \textbf{93.33} & 100 & \textbf{83.35} \\ 
\cline{2-7}
 & KNN & \textbf{83.33} & 73.33 & \textbf{93.33} & 150 & 79.98 \\ 
\cline{2-7}
 & GNB & 70.0 & 66.67 & \textbf{93.33} & 200 & 79.98 \\
\cline{2-7}
 & LR & 80.0 & 73.33 & \textbf{93.33} & 250 & 79.98 \\ 
\cline{2-7}

\thickhline
 & RF & 70.0 & 53.33 & \textbf{93.33} & 30 & \textbf{86.67} \\
\cline{2-7}
 & DT & 70.0 & 56.67 & \textbf{93.33} & 50 & 83.35 \\ 
\cline{2-7}
\textbf{SC-L2} & SVM & \textbf{83.33} & \textbf{70.0} & 90.0 & 100 & 79.98 \\ 
\cline{2-7}
 & KNN & 66.67 & \textbf{70.0} & 86.67 & 150 & 76.66 \\ 
\cline{2-7}
 & GNB & 76.67 & 66.67 & 90.0 & 200 & 60.01 \\
\cline{2-7}
 & LR & 76.67 & 60.0 & 90.0 & 250 & 63.33 \\ 
\cline{2-7}

\thickhline
 & RF & \textbf{90.0} & 53.33 & \textbf{100.0} & 30 & \textbf{83.35} \\
\cline{2-7}
 & DT & 83.33 & 53.33 & \textbf{100.0} & 50 & 76.66 \\
\cline{2-7}
\textbf{SC-L3} & SVM & 70.0 & 70.0 & 96.67 & 100 & 63.33 \\
\cline{2-7}
 & KNN & 70.0 & 50.0 & 96.67 & 150 & 60.01 \\
\cline{2-7}
 & GNB & 73.33 & 70.0 & 96.67 & 200 & 63.33 \\
\cline{2-7}
 & LR & 80.0 & \textbf{73.33} & \textbf{100.0} & 250 & 56.69 \\ 
\cline{2-7}

\thickhline
 & RF & 63.33 & 63.33 & 83.33 & 30 & 76.66 \\
\cline{2-7}
 & DT & \textbf{80.0} & 70.0 & 83.33 & 50 & \textbf{83.35} \\
\cline{2-7}
\textbf{SC-L4} & SVM & \textbf{80.0} & \textbf{73.33} & 90.0 & 100 & 79.98 \\
\cline{2-7}
 & KNN & 76.67 & 70.0 & 83.33 & 150 & 56.69 \\
\cline{2-7}
 & GNB & 76.67 & 70.0 & \textbf{93.33} & 200 & 66.65 \\
\cline{2-7}
 & LR & 76.67 & 66.67 & \textbf{93.33} & 250 & 60.01 \\
\cline{2-7}

\thickhline
 & RF & \textbf{90.0} & 66.67 & \textbf{93.33} & 30 & \textbf{79.98} \\%
\cline{2-7}
 & DT & 86.67 & 60.0 & \textbf{93.33} & 50 & 66.65 \\%
\cline{2-7}
\textbf{SC-L5} & SVM & 86.67 & \textbf{83.33} & \textbf{93.33} & 100 & 56.69 \\%
\cline{2-7}
 & KNN & 83.33 & \textbf{83.33} & \textbf{93.33} & 150 & 73.34 \\%
\cline{2-7}
 & GNB & 86.67 & 76.67 & \textbf{93.33} & 200 & 50.0 \\%
\cline{2-7}
 & LR & \textbf{90.0} & \textbf{83.33} & \textbf{93.33} & 250 & 73.34 \\%
\cline{2-7}

\thickhline
 & RF & 86.21 & 72.41 & 93.1 & 30 & \textbf{86.23} \\%
\cline{2-7}
 & DT & 86.21 & \textbf{89.66} & 93.1 & 50 & \textbf{86.23} \\%
\cline{2-7}
\textbf{6MWT$^{a}$} & SVM & \textbf{89.66} & 79.31 & \textbf{96.55} & 100 & 82.76 \\%
\cline{2-7}
 & KNN & 82.76 & 82.76 & 82.76 & 150 & 82.76 \\%
\cline{2-7}
 & GNB & \textbf{89.66} & 75.86 & 93.1 & 200 & 79.3 \\%
\cline{2-7}
 & LR & 82.76 & 86.21 & 93.1 & 250 & 79.3 \\%
\cline{2-7}

\thickhline
 & RF & 88.46 & 88.46 & 92.31 & 30 & 80.76 \\%
\cline{2-7}
 & DT & 88.46 & \textbf{92.31} & 92.31 & 50 & 80.76 \\%
\cline{2-7}
\textbf{100MRW$^{b}$} & SVM & 88.46 & 80.77 & \textbf{96.15} & 100 & 69.24 \\%
\cline{2-7}
 & KNN & \textbf{92.31} & 80.77 & 92.31 & 150 & \textbf{84.62} \\%
\cline{2-7}
 & GNB & \textbf{92.31} & 88.46 & \textbf{96.15} & 200 & 80.76 \\%
\cline{2-7}
 & LR & \textbf{92.31} & 88.46 & 92.31 & 250 & \textbf{84.62} \\%
\cline{2-7}

\thickhline
 & RF & 80.0 & 83.33 & \textbf{93.33} & 30 & 83.35 \\%
\cline{2-7}
 & DT & 80.0 & 76.67 & \textbf{93.33} & 50 & \textbf{86.67} \\%
\cline{2-7}
\textbf{FW} & SVM & \textbf{93.33} & \textbf{86.67} & 90.0 & 100 & 79.98 \\%
\cline{2-7}
 & KNN & 83.33 & 83.33 & 86.67 & 150 & 73.34 \\%
\cline{2-7}
 & GNB & 80.0 & 83.33 & 86.67 & 200 & 76.66 \\%
\cline{2-7}
 & LR & 90.0 & 80.0 & 86.67 & 250 & 76.66 \\%
\cline{2-7}
\thickhline
\end{tabular}}
\begin{flushleft}
$^{a}$ One DMD child was unable to perform the activity. \\$^{b}$ Four DMD children were unable to perform the activity. \\$^{*}$ Each second contains 100 samples.
\end{flushleft}
\end{adjustwidth}
\end{table}

CML-CF achieved the best accuracy of 100\% at SC-L3, while DL-RAW achieved 86.67\% accuracy at both SC-L2 and FW. In comparison, the CML-CF approaches utilize  CFs while DL-RAW only uses raw accelerometer data.
CFs in the CML-CF approach include features related to the activity as a whole, such as speed and step length, and participants in particular, such as age, height, and weight. We note that the DL-RAW approach does not use participant features. It also cannot extract other activity features such as speed and step length because it does not have information on the distance in each experiment. These suggest the effectiveness of CFs.

Furthermore, CML-CF considers the average metrics of each participant, while DL looks at TWs of raw signals of the activity. DL requires substantial data to train a model, so its accuracy drops with larger TW sizes in short-duration activities. The drops continue as activities get shorter with fewer TWs available for analysis (e.g., from SC-L1 to SC-L4). In more extended activities like 6MWT and FW, the differences in accuracy between TW sizes tend to decrease.

\subsubsection{CML-CF Approach}
We evaluated the CML approach using temporospatial gait CFs. We report the classifier's accuracy among 6 different CML techniques: RF, DT, SVM, KNN, GNB, and LR. For CML-CF, three different methods were used: no projection (CML), PCA projection (CML-PCA), and LDA projection (CML-LDA).  Each group of results represents one of the eight different activities mentioned previously. Our models achieved the best accuracy of 100\% in self-selected-walk speed (SC-L3) and exceeded 96\% in both fast-walk/jog/run speeds (100MRW) and fast-walk speed (6MWT).

\subsubsection{DL-RAW Approach}
We evaluated the DL approach using raw data. Since DL algorithms require a large amount of data to train, each activity was divided into fixed TWs using the window-slicing method to provide more data for the DL model. Table-\ref{table3} reports the detailed results of DL-RAW analyses and optimization of TW segment length. 

Our model demonstrated the highest accuracy of 86.67\% in slow-walk speed (SC-L2) and self-selected-walk (FW) activities. We hypothesize this was achieved using TWs long enough to capture a portion of the gait cycle with a strong correlation between the accelerometer signal's x, y, and z-axes. However, it is important to note that different portions of the gait cycle may have varying levels of distinguishability, resulting in some parts having a higher likelihood of being classified correctly than others. When the portions of gait cycles are located in a region where the correlation between the x, y, and z-axis is weak, the characteristics of those portions may not be captured effectively by the DL model. As a result, the TW may not have enough distinctive features to be classified correctly. One possibility is to use a TW that includes at least one complete gait cycle can ensure that all the portions that contain sufficient correlation between the x, y, and z axes are captured. By setting a threshold for the percentage of the contribution of each portion to the identification task, we can further enhance the accuracy and reliability of our model. This approach can potentially improve clinical explainability and enhance performance. The only limitation of this approach is that it requires more data to train the model. We plan to explore it further in future studies.

\subsection{Relationship Between Extracted Step Length, Gait Speed, and Functional Ability}
\begin{figure}[!ht]
\centerline{\includegraphics[width=1.2\textwidth]{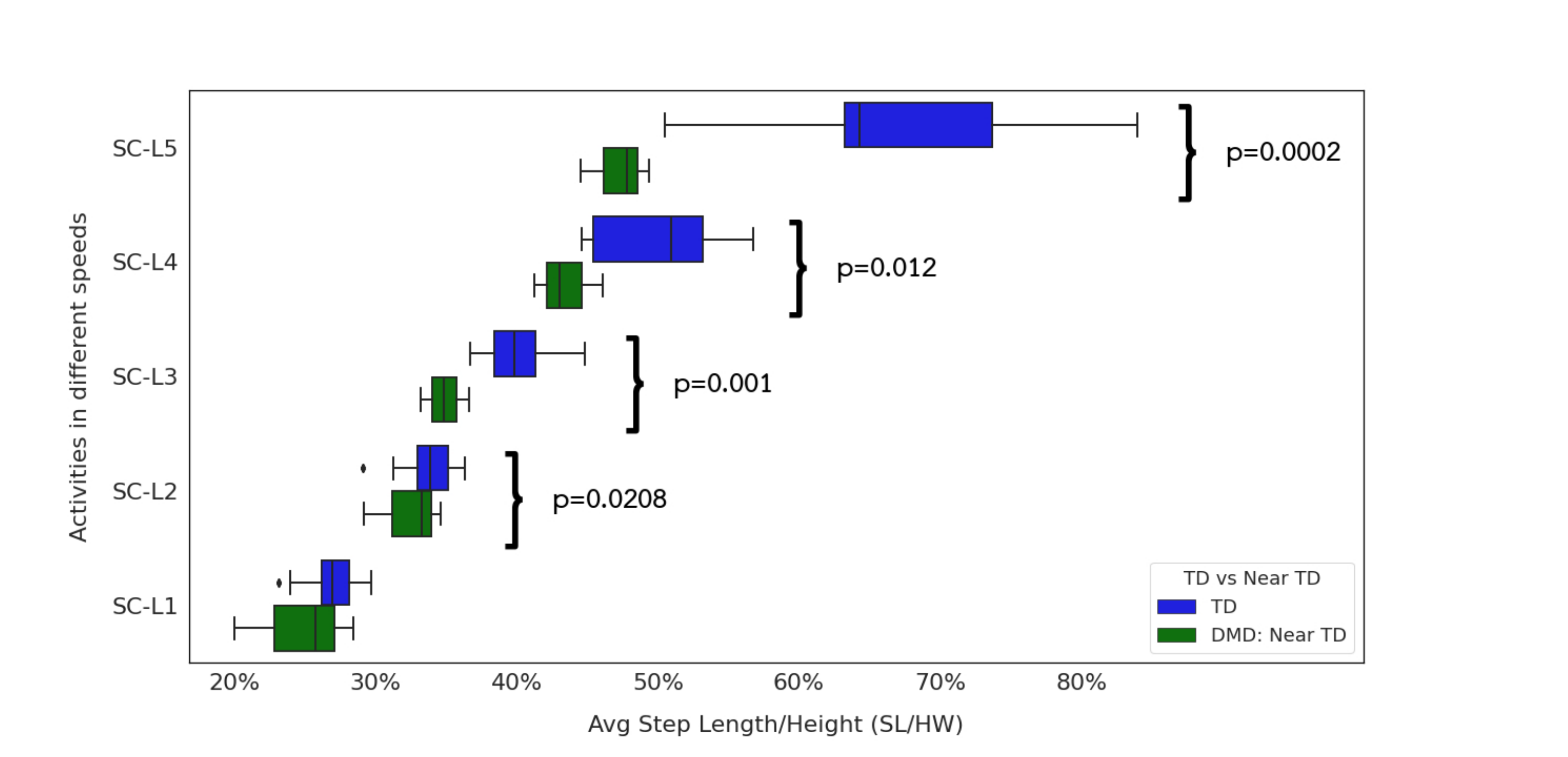}}
\caption{{\bf Step Length by Different Gait Speeds.}
TD and ambulatory DMD children with NSAA$\geq 30$.
}
\label{Fig7}
\end{figure}

\begin{figure}[!ht]
\centerline{\includegraphics[width=1.2\textwidth]{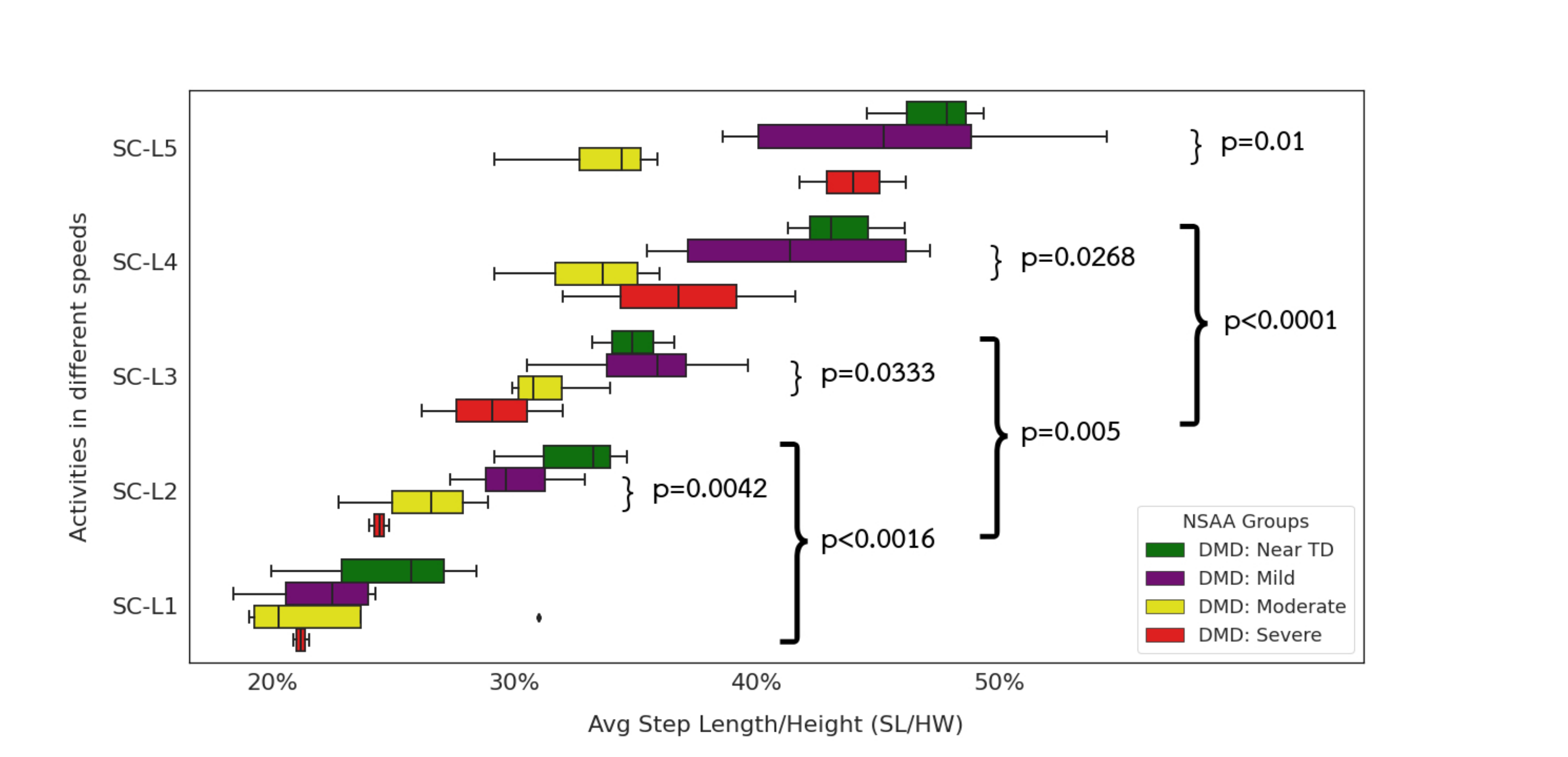}}
\caption{{\bf Step Length by Different Gait Speeds.}
NSAA severity group in ambulatory DMD children.
}
\label{Fig8}
\end{figure}

\begin{figure}[!ht]
\centerline{\includegraphics[width=1.2\textwidth]{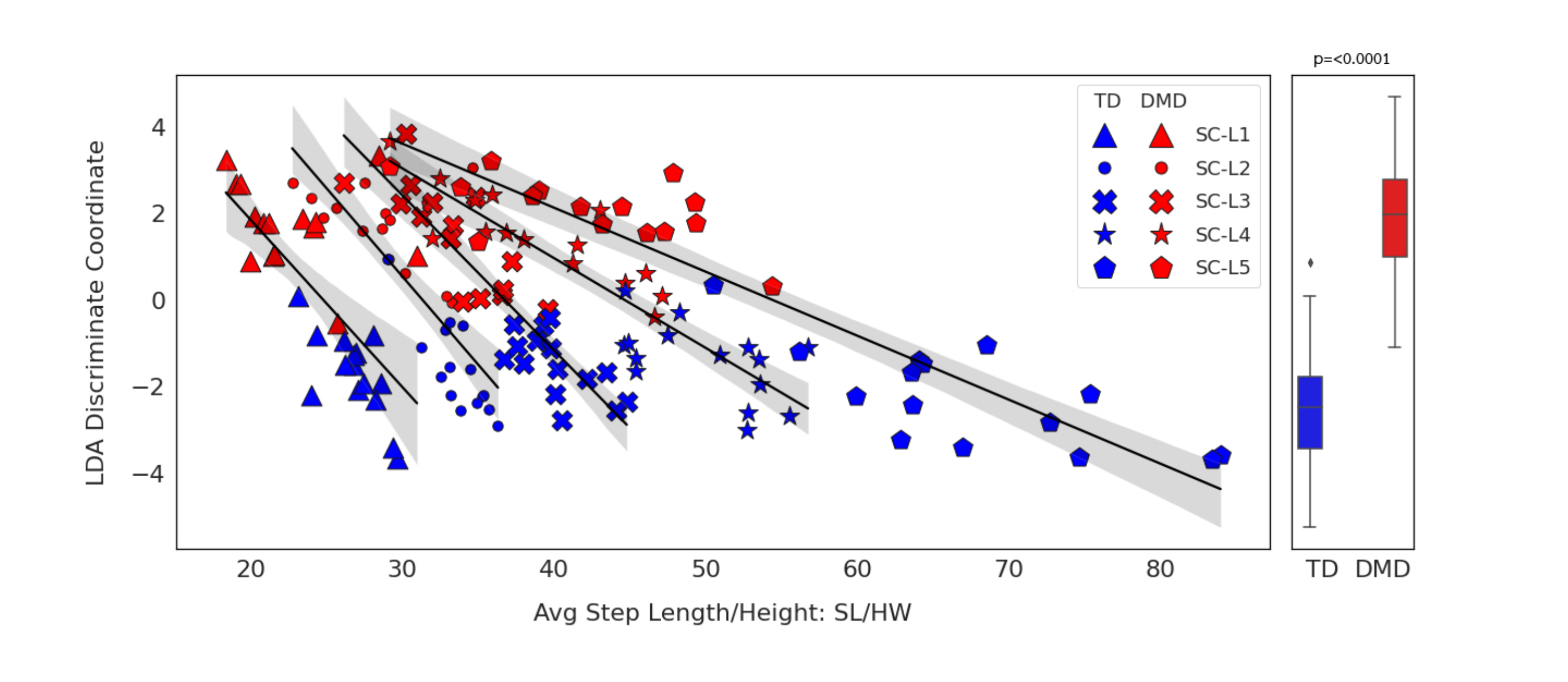}}
\caption{{\bf LDA coordinates vs. step length at different gait speeds.}
DMD vs. TD.
}
\label{Fig9}
\end{figure}

To help illustrate how step length measures relate to the overall clinical function, we compared step lengths for DMD participants with NSAA clinical function scores $\ge30$ (near-TD based on a 34-point maximum score) with TD participants at each gait speed from slow walking to jogging or running (SC-L1 to SC-L5) as shown in Fig~\ref{Fig7}. At all but the lowest slow walking speed, this highest-functioning group of DMD children demonstrated significant reductions in step length compared to their TD peers. We then subdivided all DMD participants by NSAA score into groups of near-TD ($\ge30$), mildly affected ($20-29$), moderately affected ($10-19$), and severely affected ($<10$) individuals, as shown in Fig~\ref{Fig8}. In those DMD children, average step lengths differed significantly between each velocity activity from a slow walk (SC-L1) to a fast walk (SC-L4). Furthermore, there were significant differences in step length between the mildly and moderately affected NSAA groups within the medium-slow (SC-L2), comfortable (SC-L3), and fast walk (SC-L4) activities, suggesting that the largest drop in step length appears to occur near the middle of the NSAA scoring range (scores $10-29$). We should note, however, that the largest difference in step lengths seen in the moderate NSAA group could be due to the non-linear nature of NSAA scoring.  The NSAA score is a composite of multiple ordinal item scores rather than a continuous variable.
We calculated correlations between temporospatial gait CFs and PCA and LDA dimension scores to determine which temporospatial gait CFs helped describe CML tool outputs (Table-\ref{table4}). For PCA-based analyses, the PC1 component score was significantly and highly correlated with speed, step frequency, stride length, and total power across all velocity groups. The PC1 component was also significantly and moderately correlated with NSAA clinical and functional scores. In contrast, the PC2 component score was predominantly associated with the axial percentages of total power. The unidimensional LDA score was significantly and highly correlated with step length and significantly and highly to moderately correlated with the NSAA score. The relationship between step length and the LDA coordinates at different velocities is illustrated in Fig~\ref{Fig9}. Step length increased with velocity across all velocity activities (SC-L1 to SC-L5), with positive LDA scores on average indicating DMD participants and negative LDA scores indicating TD with little overlap between groups ($p<0.0001$). 

\begin{table}[!ht]
\begin{adjustwidth}{-2.25in}{0in} 
\centering
\caption{\bf Correlation of Dimensional Reduction Scores with Extracted Clinical Features and NSAA Score}
\label{table4}
\resizebox{1.4\textwidth}{!}{
\begin{tabular}{|p{2.2cm}||p{1.6cm}|p{1.6cm}||p{1.45cm}|p{1.6cm}|p{1.45cm}|p{1.45cm}|p{1.45cm}|p{1.45cm}|p{1.45cm}|p{1.45cm}||p{1.45cm}|}  
\thickhline
  \multicolumn{1}{|c||}{\textbf{}} 
& \multicolumn{1}{c|}{\textbf{}} 
& \multicolumn{1}{c||}{\textbf{}} 
& \multicolumn{8}{c||}{\textbf{Temporospatial Gait Clinical Features (CFs)}} 
& \multicolumn{1}{c|}{\textbf{}} \\
\cline{4-11}
\textbf{Components} & \textbf{Activities} & \textbf{Value} & \textbf{SP} & \textbf{SF} & \textbf{SL} & \textbf{TP} & \textbf{VP} & \textbf{MP} & \textbf{AP} & \textbf{FI} & \textbf{NSAA$^{*}$} \\ 
\thickhline

& SC-L1& Correlation\newline(p-value)& 0.93\newline(\textbf{$<$0.0001})& 0.85\newline(\textbf{$<$0.0001})& 0.72\newline(\textbf{0.0002})& 0.82\newline(\textbf{$<$0.0001})& 0.03\newline(1.0)& -0.44\newline(0.5332)& 0.55\newline(0.0589)& 0.81\newline(\textbf{$<$0.0001})& 0.56\newline(\textbf{0.0014}) \\ \cline{2-12}
& SC-L2& Correlation\newline(p-value)& 0.92\newline(\textbf{$<$0.0001})& 0.85\newline(\textbf{$<$0.0001})& 0.61\newline(\textbf{0.0116})& 0.83\newline(\textbf{$<$0.0001})& -0.28\newline(1.0)& -0.32\newline(1.0)& 0.74\newline(\textbf{0.0001})& 0.74\newline(\textbf{0.0001})& 0.4\newline(\textbf{0.0273}) \\ \cline{2-12}
\textbf{ PC1$^{a}$ }& SC-L3& Correlation\newline(p-value)& 0.93\newline(\textbf{$<$0.0001})& 0.85\newline(\textbf{$<$0.0001})& 0.54\newline(0.0679)& 0.83\newline(\textbf{$<$0.0001})& 0.29\newline(1.0)& -0.41\newline(0.8368)& 0.17\newline(1.0)& 0.72\newline(\textbf{0.0003})& 0.51\newline(\textbf{0.0041}) \\ \cline{2-12}
& SC-L4& Correlation\newline(p-value)& 0.91\newline(\textbf{$<$0.0001})& 0.85\newline(\textbf{$<$0.0001})& 0.66\newline(\textbf{0.0027})& 0.83\newline(\textbf{$<$0.0001})& 0.82\newline(\textbf{$<$0.0001})& -0.78\newline(\textbf{$<$0.0001})& -0.23\newline(1.0)& 0.7\newline(\textbf{0.0007})& 0.71\newline(\textbf{$<$0.0001}) \\ \cline{2-12}
& SC-L5& Correlation\newline(p-value)& -0.93\newline(\textbf{$<$0.0001})& -0.85\newline(\textbf{$<$0.0001})& -0.83\newline(\textbf{$<$0.0001})& -0.89\newline(\textbf{$<$0.0001})& -0.66\newline(\textbf{0.0023})& 0.88\newline(\textbf{$<$0.0001})& -0.2\newline(1.0)& -0.76\newline(\textbf{$<$0.0001})& -0.62\newline(\textbf{0.0002}) \\ \cline{2-12}
\thickhline
& SC-L1& Correlation\newline(p-value)& 0.12\newline(1.0)& 0.15\newline(1.0)& 0.01\newline(1.0)& 0.35\newline(1.0)& -0.76\newline(\textbf{0.0001})& 0.89\newline(\textbf{$<$0.0001})& -0.52\newline(0.1481)& 0.14\newline(1.0)& -0.28\newline(0.1278) \\ \cline{2-12}
& SC-L2& Correlation\newline(p-value)& 0.19\newline(1.0)& 0.11\newline(1.0)& 0.2\newline(1.0)& -0.43\newline(0.7552)& 0.85\newline(\textbf{$<$0.0001})& -0.9\newline(\textbf{$<$0.0001})& 0.19\newline(1.0)& -0.4\newline(1.0)& 0.19\newline(0.3069) \\ \cline{2-12}
\textbf{ PC2$^{b}$}& SC-L3& Correlation\newline(p-value)& 0.19\newline(1.0)& -0.32\newline(1.0)& 0.57\newline(\textbf{0.0485})& -0.53\newline(0.1187)& 0.85\newline(\textbf{$<$0.0001})& -0.72\newline(\textbf{0.0003})& -0.25\newline(1.0)& -0.57\newline(\textbf{0.0453})& 0.3\newline(0.1049) \\ \cline{2-12}
& SC-L4& Correlation\newline(p-value)& -0.19\newline(1.0)& 0.3\newline(1.0)& -0.61\newline(\textbf{0.0144})& 0.27\newline(1.0)& -0.12\newline(1.0)& 0.45\newline(0.603)& -0.65\newline(\textbf{0.004})& 0.29\newline(1.0)& -0.41\newline(\textbf{0.0253}) \\ \cline{2-12}
& SC-L5& Correlation\newline(p-value)& -0.09\newline(1.0)& -0.17\newline(1.0)& 0.05\newline(1.0)& -0.42\newline(0.9396)& 0.56\newline(0.0645)& 0.11\newline(1.0)& -0.93\newline(\textbf{$<$0.0001})& -0.54\newline(0.0945)& 0.29\newline(0.1268) \\ \cline{2-12}
\thickhline
& SC-L1& Correlation\newline(p-value)& -0.54\newline(0.0695)& -0.35\newline(1.0)& -0.7\newline(\textbf{0.0007})& -0.04\newline(1.0)& -0.07\newline(1.0)& 0.47\newline(0.3336)& -0.55\newline(0.0601)& -0.42\newline(0.7212)& -0.72\newline(\textbf{$<$0.0001}) \\ \cline{2-12}
& SC-L2& Correlation\newline(p-value)& -0.51\newline(0.1305)& -0.11\newline(1.0)& -0.78\newline(\textbf{$<$0.0001})& 0.05\newline(1.0)& -0.26\newline(1.0)& 0.42\newline(0.7424)& -0.25\newline(1.0)& -0.3\newline(1.0)& -0.63\newline(\textbf{0.0002}) \\ \cline{2-12}
\textbf{ LDA }& SC-L3& Correlation\newline(p-value)& -0.64\newline(\textbf{0.0054})& -0.02\newline(1.0)& -0.91\newline(\textbf{$<$0.0001})& -0.05\newline(1.0)& -0.46\newline(0.3801)& 0.56\newline(\textbf{0.044})& -0.13\newline(1.0)& -0.23\newline(1.0)& -0.68\newline(\textbf{$<$0.0001}) \\ \cline{2-12}
& SC-L4& Correlation\newline(p-value)& -0.65\newline(\textbf{0.004})& -0.19\newline(1.0)& -0.89\newline(\textbf{$<$0.0001})& -0.01\newline(1.0)& -0.54\newline(0.0732)& 0.54\newline(0.0795)& 0.1\newline(1.0)& -0.07\newline(1.0)& -0.79\newline(\textbf{$<$0.0001}) \\ \cline{2-12}
& SC-L5& Correlation\newline(p-value)& -0.87\newline(\textbf{$<$0.0001})& -0.59\newline(\textbf{0.0195})& -0.92\newline(\textbf{$<$0.0001})& -0.42\newline(0.8081)& -0.59\newline(\textbf{0.0204})& 0.61\newline(\textbf{0.0134})& 0.05\newline(1.0)& -0.18\newline(1.0)& -0.67\newline(\textbf{0.0001}) \\ \cline{2-12}
\thickhline

\end{tabular}}
\begin{flushleft}
$^{*}$ The Spearman rank-order correlation coefficient. The level of significance was set at 0.05.\\
$^{a}$ Captures the most variation.\\
$^{b}$ Captures the second most variation.
\end{flushleft}
\end{adjustwidth}
\end{table}

\section{Discussions}
The purpose of our study was to explore the utility and feasibility of collecting clinically meaningful gait data using consumer-level accelerometers outside of the formal gait laboratory setting and to explore a range of CML and DL methods to differentiate between children with DMD and TD of different ages. Extracting and describing a combination of well-known and understood temporospatial gait CFs allowed us to identify some of the characteristics that CML and DL tools used to differentiate between DMD and TD groups.    
\subsection{Utility and Feasibility of CML and DL Approaches to Extracted Temporospatial Gait CFs}
We investigated two different ML approaches, CML with extracted temporospatial gait CFs and DL with RAW data. We reported the outcome for each gait velocity in Table-\ref{table3}.  We also showed that using CML with extracted temporospatial gait CFs to predict membership in the DMD group yielded satisfactory results, with correct predictions for up to 100\% of participants. The CML-CF approach typically shows improved accuracy with gait CFs, and we thus expect that identifying and extracting more correlated temporospatial gait CFs will improve the current model's outcome in future studies. This improvement in the outcome depended on how these new temporospatial gait CFs related to the gait cycle quantitatively.

In comparison, DL has been shown to have a high accuracy in several medical fields~\cite{REF_REVIEW}. It also does not require feature engineering as in CML. At the same time, DL requires a large amount of data to train and DL models lack explainability, which might be concerning in medical fields.  However, our promising results comparing DL-RAW with CML-CF approaches should encourage researchers to conduct further research so that we can transfer the knowledge yielded from DL-RAW to improve the existing temporospatial gait CFs and, and conversely,  use temporospatial gait CFs to aid in interpreting DL-RAW results.

\subsection{CML-CF Approach }
\subsubsection{Extracted Gait Features are Consistent with Clinical Observations}
It is commonly known that in people with DMD, the temporospatial gait characteristics of speed, step frequency, and step length are on average lower than those in TD peers. We extracted temporospatial gait CFs from signals derived from a single mobile phone-based triaxial accelerometer using methods similar to those described by Barthelemy~\cite{dog}, and our data demonstrated that across a range of commonly-attained speeds, our extracted gait features differ between DMD and TD controls with many differences reaching statistical significance. 

Analysis of power spectra of time-series gait data using single sensors during walking has been explored in comparison to some mobility-limited human populations but has not, to our knowledge, been previously applied in a DMD population.  The series of papers by Barthelemy~\cite{dog}~\cite{ dogPCA} using single sensors to measure gait characteristics of dystrophic GRMD dogs and to evaluate extracted temporospatial gait CFs using LDA methods demonstrated the utility of such methods and provided inspiration for our approach. The overall decreased vertical percent power and increased lateral power that we observed appear consistent with clinical observations of the development of a lateral, trendelenburg gait pattern in children with DMD as described by D’Angelo~\cite{REF_2}. In TD individuals, as walking speed increases and progresses to running, ground clearance for the foot and leg in the swing phase is achieved through a gradual but proportional increase in vertical movement. In people with DMD with progressing weakness, however, swing phase leg clearance is achieved through substitution using a more pronounced lateral shift of the center of mass to the stance phase leg with the elevation of the contralateral hip. This more lateral gait style is effective but also less efficient, and results in greater work for reduced forward motion as demonstrated by prior studies showing increased heart rate-based energy expenditure of ambulation with increasing step frequency in DMD children measured by COSMED portable metabolic testing combined with StepWatch activity monitoring~\cite{mcdonald2005use}.

\subsubsection{Interpreting extracted clinical features}
A major challenge with the use of CML approach in evaluating health status is to interpret model outputs relative to well-known temporospatial gait CFs of a disease. Here we have built on our colleague’s work in GRMD dogs to extract similar clinically-salient temporospatial CFs of gait from our accelerometer data in humans to help explain differences between our DMD and TD cohorts. In addition, we evaluated our participants using common quantitative timed motor performance tests (25 meters, 100 meters fast-walk/jog/run tests, and the 6MWT) and the NSAA, which all have demonstrated utility as outcome measures used in clinical care and clinical trial contexts.

We do not show correlations between timed motor performance measures and extracted clinical features because the features themselves (velocities, step lengths, and step frequencies) are derived directly from the test’s times and distances (Table-\ref{table4}). NSAA, however, is not a feature in our models and serves as an external “anchor” to help with the overall interpretation of model performance and to provide additional cross-validation of step length.

\begin{figure}[!ht]
\centerline{\includegraphics[width=1.2\textwidth]{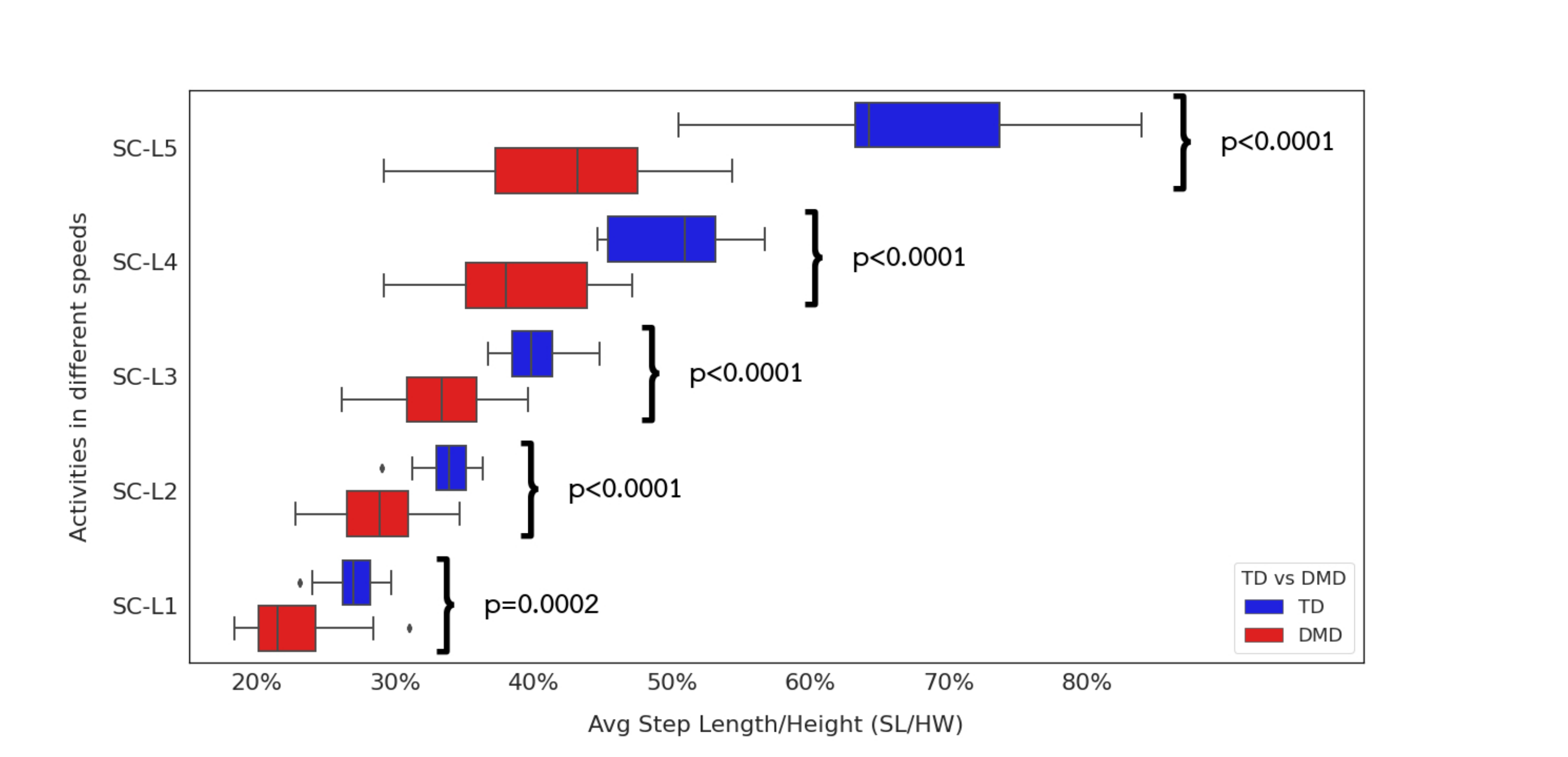}}
\caption{{\bf Step Length by Different Gait Speeds.}
TD vs. DMD children. This boxplot displays the data distribution based on: minimum, first quartile (25th percentile), median, third quartile (75th percentile), and maximum, as well as the outliers.
}
\label{Fig10}
\end{figure}

At near-TD levels (NSAA $\geq 30$) near the ceiling (Fig~\ref{Fig7}), children with DMD show reduced height-normalized step length compared to TD peers at all but the slowest of walks. The most dramatic differences occur in the fast walking and jogging or running paces. Looking in more detail at the DMD children (Fig~\ref{Fig8}), we can see significant differences in all gait paces except in SC-L1 pace.  At the same time, we fail to see any significant differences between step lengths for fast walks and jogging or running paces. This supports the clinical observation that some children with DMD fail to develop a “double-off” running pattern that allows TD individuals to lengthen their step lengths well past what they can achieve at a fast walk (Fig~\ref{Fig10}).

\subsubsection{Utility of CML Models as Classifiers}
CML with LDA dimensionality reduction (CML-LDA) surpasses both the original CML (without dimensionality reduction) and CML-PCA in the self-selected-walk to run speeds. For example, CML-LDA achieves a maximum accuracy of 100\% in SC-L3 and exceeds 96\% in 6MWT and 100MRW. 

Our results indicate that gait speed is an essential component in single accelerometer gait analysis~\cite{REF_4} and that changes in gait speed can affect classification accuracy. We believe the observation that speed and accuracy are correlated would encourage researchers to consider gait speed's effect on gait characteristics. In this respect, CML approaches seem to rely on elements that are highly affected by gait speed, e.g., the observed differences in lateral movement or shortened relative step length that are apparent to the eye and thus picked up by domain experts. It would be interesting to further investigate if additional CFs could be developed and used to improve classification accuracy. 

\subsubsection{CML with Dimensional Reduction}
Using PCA and LDA provides a 1D and 2D representation of the DMD and TD group's distribution. We reduce the dimensionality of the temporospatial gait CFs to 2D using PCA and 1D using LDA. Both TD and DMD participants’ temporospatial gait CFs after PCA and LDA reduction tend to form different groups. LDA (supervised) maximizes the separability between classes, and PCA (unsupervised) maximizes the variance within the classes. The resulting visual representation of these group separation methods provides valuable feedback about model performance as well as a degree of difference between groups, even in the presence of complex, multi-dimensional data. When the distance between the groups becomes more obvious, the two groups tend to be separable. This gives an indicator that the model would yield a high classification accuracy. On the other hand, when the two groups intersect, which indicates high similarity between the two groups, DMD and TD, the classifier yields low accuracy. In CML-LDA (Table-\ref{table3}), we notice that the model's accuracy is best in the self-selected-walk to fast-walk speeds (SC-L3 and 6MWT), and fast-walk/jog/run speed (100MRW). We can see that the accuracy of the CML-LDA overall performs better than both the original CML and CML-PCA models. In 6MWT and 100MRW activities, we observe that both the original CML and CML-PCA reach their highest accuracy of 89.66\% and 92.31\%, respectively. Therefore, CML-LDA is more accurate and provides a helpful visual representation in the self-selected-walk to fast-walk speeds (SC-L3 and 6MWT), and fast-walk/jog/run speed (100MRW) activities which aids to better understand the differences in gait patterns of the participants. 

\subsubsection{Correlation of PCA and LDA models with clinical features}
PCA-based classification tools perform slightly worse than LDA-based approaches. However, the 2-major components of PCA results (PC1 and PC2) are interesting in that the first factor correlates well with temporospatial gait CFs of height-adjusted step velocity, step frequency, and total power. The other one correlates more strongly with proportional axial portions of power in the vertical, mediolateral, and anteroposterior directions. Viewed from the perspective of progressive DMD symptoms with increased weakness, shortening overall step lengths, and more lateral gait, the selection of these two sets of features as drivers of differences between DMD and control groups makes clinical sense. The LDA-based approaches, however, appear more accurate in differentiating between DMD and TD children. From the perspective of the correlation of LDA location scores and clinical features, it is not surprising that the model outputs correlate most with height-normalized step length as an indicator of typical walking patterns versus those affected by the DMD diagnosis (Table-\ref{table4}) and that the models are able to differentiate between groups across a broad range of self-selected velocities (Fig~\ref{Fig9}).

\subsubsection{Impact of reduced stride length on community mobility}
Besides differentiating patients with DMD from similarly-aged TD children, reductions in stride length are interesting for another reason. Children with DMD with mild to moderate gait disturbance can take similar numbers of steps during most days relative to TD peers – a key reason that step-counting devices have been less informative as outcome measures for clinical trials~\cite{lott2021walking}. However, the incorporation of distance traveled (whether adjusted to standing height or not) may provide us with valuable additional information about the effect of step length reduction on overall community travel. For instance, a recent meta-analysis by Conger et al.~\cite{conger2022time} suggests an expected average daily step count of approximately 11,000 steps in TD children (mixed male/female). If we assume that a 50th percentile height American 9-year-old is 128 centimeters tall~\cite{PMID12043359}, and that their activity is primarily a comfortable self-selected-walk, we can expect based on our control data for step length to be approximately 40\% of their standing height, or 51.2 centimeters, and that their total daily distance traveled would be 5,632 meters. A similarly-sized mildly-affected ambulatory child with DMD would have a step length of approximately 35\% of their standing height, or 44.8 centimeters and their total distance traveled would be 4,928 meters, an overall reduction of 704 meters (12\%) traveled per day compared to their TD peer. If that same DMD child at a future date was more severely affected because of disease progression, holding their step count and height equal, their step length might be 30\% of their standing height, or 38.4 centimeters and their total distance traveled would be 4,224 meters, an overall reduction of 1,408 meters (25\%) per day compared to their typical peer and 704 meters (14.2\%) less compared to their own prior performance. Granted, this brief exercise makes some unlikely assumptions, including that step counts would match, that all travel would be at a comfortable self-selected pace, that our DMD child did not grow between testing events. Clinical observations would tell us that there is also a progressive impact of fatigue that reduces overall step counts and active time~\cite{mcdonald2005use}, and that linear growth is often reduced in children with DMD both due to disease and due to the effects of steroid treatments. These factors might further reduce daily travel, or they might cancel each other out. Further research on community travel patterns and behaviors in people with DMD will be required if these questions are to be answered to most readers’ satisfaction, but it is not difficult to imagine that reduced stride lengths could noticeably limit a person’s community-level travel and participation in daily activities.

\subsection{DL-RAW Approach }
\subsubsection{Evaluation of Raw Data Using DL Models}
DL models (supervised in our study) depend heavily on input data. When we apply the DL approach to raw accelerometer data (DL-RAW), the model achieves a maximum accuracy of 86.67\% in slow-walk speed (SC-L2) and self-selected-walk speed (FW). The advantage of using DL-RAW is this method does not need feature engineering. We believe a further investigation into improving the ML algorithms that target raw data could be a promising direction. Still, one challenge is that they are still more difficult to explain from a clinical perspective than methods anchored to clinical features and evaluations.
Therefore, when the clinical explanation is a primary focus, it is favorable to extract temporospatial gait CFs that could help understand how the decisions are made in ML models. Our results indicate that gait speed is an essential component in single accelerometer gait analysis~\cite{REF_4} and that changes in gait speed can affect classification accuracy.

\subsubsection{Time-windowing of Raw Data}
By examining different TW sizes, we found that the optimal TW size should be long enough to include a portion of the gait cycle where contrasts are expected between DMD and TD children. The model determines per each single TW whether it belongs to DMD or TD children by identifying the difference in gait patterns among the DMD and TD groups. This allows us to simplify our methods and use the DL method on minimally processed raw data in a manner that requires less expertise to extract temporospatial gait CFs. Having enough samples in each TW ensures a solid correlation between the x,y, and z-axis and helps the model correctly classify. Using a small TW that does not have enough correlation between the three axes could lead to insufficient data that cannot capture gait patterns.
On the other hand, using a large TW results in a smaller number of TWs per participant. At different velocities, people with DMD and TD have similarities in some portions of their gait cycles, and even at faster velocities that are more difficult for people with DMD to achieve, some portions of gait cycles may appear more typical than others. By examining the typical/atypical decisions for each TW as mentioned in section \ref{sec:TW}, it may be possible to use that percentage of typical to atypical portion of each gait cycle across a range of speeds to indicate the severity of disease at a given point in time and to track and quantify disease progress over time.

\subsection{Effectiveness of CML and DL Models Differs Depending on Gait Velocity and Type of Gait}  
Using the CML-CF approach performs better than the DL-RAW. Overall, DL-RAW accuracy increases with the length of the activity. Since SC-L2 is a slow-walk and long-period activity compared with SC-L3 to SC-L5, DL-RAW achieves an accuracy of 86.67\%. In FW, where participants walk as fast as possible for a long duration, DL-RAW in this activity achieves an accuracy of 86.67\% while CML-CF exceeds an accuracy of 93\%. Additionally, the running efforts on the 25 meters course (SC-L5) are of short duration, which further reduces available training data. In 100MRW, where the duration of the activity is longer and provides more data to the model, the model achieves an accuracy of 84.62\% despite the fact that only some participants run while others walk.

\subsection*{Study Limitations}
The main limitation of this study is the small number of participants. Our ongoing work aims to expand the scope of this experiment to a larger group of participants spanning multiple diagnoses and age groups. An additional limitation is the short-term nature of our data collection. Multi-day data will be required to demonstrate an actual reduction in travel distance or the ability to differentiate between DMD and typical-developing groups in free-living conditions.

\section{Conclusion}
Use of ubiquitous and widely available mobile devices with single accelerometers to remotely measure differences in common clinical gait parameters represents an opportunity to expand the study of temporospatial gait characteristics into the community setting.  Our initial laboratory-based studies demonstrate ability to measure selected gait parameters across a range of typical ambulatory velocities in DMD and TD children to detect significant differences in temporospatial gait CFs that are consistent with previous studies, as well as to detect differences in proportional power of accelerations in vertical, mediolateral and anteroposterior axes. By using these clinical gait parameters and raw data and employing both CML and DL models, we are able to correctly predict whether sensor data is derived from children with DMD up to 100\% of the time at self-selected-walk pace (SC-L3). By combining these approaches, we anticipate that through ongoing studies, we will be able to improve predictive accuracy and identify additional clinically-useful parameters indicating typical growth and development, gait impairment and disease progression across a wide range of individuals with neuromuscular disease.  

\section*{Data Availability}
The authors commit to providing data access in compliance with PLOS ONE journal, grant sponsor, and University of California guidelines. Requests for data access can be addressed to the corresponding author.

\section*{Funding Acknowledgement}
This study was partially funded for participant assessment and data collection by a grant from the U.S. Department of Defense (W81XWH-17-1-0477), a pilot grant from the University of California Center for Information Technology Research in the Interest of Society (CITRIS) and the Banatao Institute, and by a research grant from the Muscular Dystrophy Association.

\section*{Acknowledgments}
We thank Zainul Abi Din, Mohammad Newaz Sharif, Vehbi Esref Bayraktar, and Ammar Haydari for their diligent proofreading. We would also like to thank Erica Goude and Omaid Sarwary for their assistance in project management. We would like to thank all of our participants and families for donating their time to this project.

\section*{Declaration of Interest}
The authors declare that they have no competing interests and no conflicts to declare.

\section*{CRediT Statement}
Albara Ah Ramli, Ph.D. candidate: Conceptualization, Methodology, Software, Formal analysis, Writing, Supervision, Validation, Visualization, Investigation, Data Curation. Xin Liu, Ph.D.: Writing - Review and Editing, Methodology, Supervision. Kelly Berndt: Investigation, Data Curation, Writing - Review and Editing. Erica Goude, MS: Investigation, Supervision, Writing - Review and Editing.  Jiahui Hou, Ph.D.: Writing - Review and Editing. Lynea  B. Kaethler, MS: Investigation, Data Curation, Writing - Review and Editing. Rex Liu, Ph.D. candidate: Writing - Review and Editing. Amanda Lopez, MS: Investigation, Data Curation, Writing - Review and Editing. Alina Nicorici, BS: Investigation, Data Curation, Methodology. Corey Owens, MS: Investigation, Data Curation. David Rodriguez, BS: Investigation, Data Curation, Writing - Review and Editing. Jane Wang: Investigation, Data Curation, Writing - Review and Editing. Huanle Zhang, Ph.D.: Writing - Review and Editing. Daniel Aranki, Ph.D.: Conceptualization, Methodology, Software, Analysis.  Craig M. McDonald, MD: Conceptualization, Resources, Funding acquisition. Erik K. Henricson, Ph.D., MPH: Conceptualization, Methodology, Software, Formal analysis, Writing, Supervision, Funding acquisition, Investigation.

\nolinenumbers

%
%
%

\bibliography{cas-refs}

\end{document}